\documentclass[]{aa}
\usepackage[]{natbib}
\usepackage{graphicx}
\usepackage{longtable}
\usepackage{lscape}
\begin{document}
%
   
	\title{A multiwavelength study of the IRAS Deep Survey galaxy sample}
   
   \subtitle{III. Spectral classification and dynamical properties}

   \author{D. Bettoni
          \inst{1}
          \and
          P. Mazzei
          \inst{1}
          \and
	   A. della Valle
	  \inst{1, 2}
	  }

   \offprints{D. Bettoni}

   \institute{INAF - Osservatorio Astronomico di Padova,
              Vicolo dell'Osservatorio, 5 -35122 Padova-ITALY\\
              \email{daniela.bettoni@oapd.inaf.it, paola.mazzei@oapd.inaf.it}
         \and
            INAF - Osservatorio Astronomico di Bologna,
          Via Ranzani, 1 - 40127 Bologna-ITALY\\
              \email{adevalle@gmail.com}
             }

   \date{}

   \abstract
{
The infrared deep sample (IDS), in the north ecliptical polar region (NEPR), is the first
complete, far--IR selected sample, on which numerous studies of
galaxy evolution are based. Such a sample allows direct investigation of the evolution of
dusty galaxies up to a redshift of about 0.3, where the global star formation rate  is known
to evolve very fast. As discussed in previous papers, we performed optical and IR (ISOCAM, 15\,$\mu$m,) follow-up
of its galaxies and exploited our IR observations to correct the 60\,$\mu$m fluxes for confusional effects and observational biases. In them we found indications of a significant incompleteness of IDS sample below  S(60)$\simeq$80\.mJy.  We constructed 15\,$\mu$m and far-IR (60\,$\mu$m)  luminosity functions of a complete sample of 56 ISO/IRAS sources. }
{Here we present and analyze  the spectral classification of several galaxies in the IDS sample  together with rotation curves which allow estimating the lower mass limits of a subsample of objects.}
{We measured fluxes and intensity ratios
of the emission lines in the visible region of the spectrum ($\lambda 4000-9000 \AA$)  for 75
galaxy members. Moreover,  for some of them (55\%), the spectra obtained  with
 the Keck II telescope have sufficient wavelength and spatial
resolution to derive their rotation curve.}
{These galaxies turn out to be disk like systems, with a  high fraction ($\sim50\%$) of interacting systems.
The spectroscopic classification of 42 galaxies, using  the emission-line ratio diagnostic diagrams, shows that the NEPR sample is predominantly composed of starburst
galaxies (71\%), while the fraction of AGNs (7\%) and LINERs (21\%)  is small. The dynamical analysis allows us to estimate the lower mass limits of 39 galaxies.}
{The rest-frame FIR luminosity distribution of these galaxies  spans the same range as that of the FIR selected complete sample, i.e. three orders of magnitude, with the same mean value, log($L_{FIR}$)=10.2. This emphasizes that such galaxies represent FIR properties of the whole sample well. Moreover, their optical properties are typical of the sample itself since 62\%  of these  belong to the 60\,$\mu$m selected complete sample.}


\authorrunning{Bettoni et al.}
 
	\titlerunning{Multi--wavelength study of the IDS/ISOCAM sample.III}

   \maketitle

 \maketitle
%
\defcitealias{DV05}{Paper~I}
\defcitealias{Maz07}{Paper~II}
\section{Introduction}

One of the most useful tools for studying the physical conditions in galactic nuclei
is the analysis of the interstellar medium. The warm ionized gas ($\sim 10^4$ K),
for example, is present in the nuclei of nearby galaxies of all morphological types.
Detected through optical emission lines, this component of the 
interstellar medium has served as a diagnostic of the physical conditions in 
galactic nuclei \citep{hfs93}, including their excitation source and chemical 
abundances, and it allows spectral classification to define the star-forming LINERS and AGN regions.
A variety of optical emission-line ratio diagnostics have been presented and employed to determine metallicities, abundances and star formation rates (SFR). In particular, using [NII]$\lambda$6583/H$\alpha$ - [OIII]$\lambda$5007/H$\beta$ and  [SII]$\lambda$6717/H$\alpha$ - [OIII]$\lambda$5007/H$\beta$ ratios diagnostics  brings out the separations among the three classes \citep{bpt81,vo87}. The separation of extragalactic objects according to their primary excitation mechanisms, together with their photometric and  kinematical properties, leads to a more complete view of their structure and evolution.
The spatially extended resolved emission lines provide powerful tracers of the  kinematics of a galaxy. These kinematical data are very important because they allow us to derive dynamical masses and to compare them with those obtained from models. In this view, the study of the dynamics of dusty galaxies in the very local universe gives more insight into their evolution.
 
Far-IR galaxies are dusty objects where a mixture of ongoing AGN activity and star formation play their roles. They are mostly interacting and merging gas-rich spirals \citep{LeFlo05} and may represent a link in an evolutionary sequence from spiral galaxies to ellipticals, via mergers. With this connection to early galaxy formation and evolution, the local population can be thought of as a laboratory for studying processes in detail such as SF triggering and starburst vs. AGN interplay. Recent studies suggest many AGNs in IR luminous and massive galaxies \citep{Treisteretal10}, and \citet{Melb08} 
have shown that most of dusty galaxies at z$\sim$0.6 are quite regular disks without evidence of strong interactions.

For this purpose the IRAS Deep Survey (IDS) sample, defined by the \citet{HH87} co-adding IRAS scans of the north ecliptic polar region and representing more 
than 20 hours of integration time, is one of the best-suited samples. \citet{Maetal01} exploited  ISOCAM observations (range 12-18 $\mu$m) of 94 IRAS Deep Survey (IDS) fields \citep{Aus00}, 
centered on the nominal positions of IDS sources, to correct the 60\,$\mu$m fluxes for confusion effects and observational biases. They found indications of a significant incompleteness of the IDS sample below S(60)$\simeq 80\,$mJy.
In \citet{DV05}, the first paper of this series (\citetalias{DV05} in the following), we presented spectroscopic and optical observations of candidate identifications  of our ISOCAM sources, and the redshift distribution of the 60\,$\mu$m  complete subsample defined by \citet{Maetal01} comprising 56 sources.
In the
second paper,  \citet{Maz07} (\citetalias{Maz07} in the following), we derived  the 60\, $\mu$m luminosity function (LF) and
the poorly known  15\, $\mu$m LF with the bivariate method of such a sample.

The optical data, together with IR and far-IR (FIR) ones, are very useful for an integral view of the properties of galaxies in our sample. In particular we will derive the spectral energy distribution (SED) extended over several orders of magnitude in wavelength to analyze the evolution of these dusty galaxies over several Gyr in look-back time, i.e., over an interval of time in which the average SFR in the Universe is known to evolve strongly (in prep).

In this paper we give new spectroscopic observations of  75 galaxies and classification of 42 galaxies, thus extending our previous far-IR (FIR) data to the optical, an essential step, in getting a complete view of these galaxy properties.  Moreover, dynamical parameters are derived for 41 galaxies in the sample.
 The plan of the paper is the following. In Sections 2 we summarize our
spectroscopy runs, which are fully described in \citetalias{DV05}, and present the more important spectral corrections. In Section 3 we analyze the spectral properties of our sample inside a fixed physical radius of  3\,Kpc, to derive their 
general  classification. Section 4 focuses on the analysis of dynamical parameters and the rotation curves available for 41 and 31 of them, respectively. This analysis allowed us to derive mass lower limits for 39 galaxies in the sample. Finally Section 5 gives our conclusions. Here we adopt $\Lambda$=0.7, $\Omega_0$=0.3 and $H_0$=70 km\,s$^{-1}$\,Mpc$^{-1}$.

\begin{table*}
\caption{Instrument set-up for every night of observations (see also \citealt{DV05})}
\label{log_obs}
\begin{tabular}{lcclcccc}
\hline\hline
Telescope   & Run       & ($''$)/pix & Grism  & Slit  & $\lambda$ & Res\\
& & & grating & ($''$) & range(\AA)& km/sec\\
\hline
Ekar+AFOSC  & 28/6/00   & 0.473     & \#4      & 2.1  & 3360-7740 & 470 \\
TNG+DOLORES & 19/5/01   & 0.275     & LR-R     & 1.5  & 4470-10360 & 300 \\
Keck II+ESI & 17/7/01   & 0.168-0.120     & 175 lines mm$^{-1}$ & 1.0  & 3900-11000 & 35 \\
Ekar+AFOSC  & 25/8/01   & 0.473     & \#4, \#8 & 2.1  & 3360-7740 & 470 \\
Ekar+AFOSC  & 12/9/01   & 0.473     & \#8      & 2.1  & 6250-8050 & 190 \\
Ekar+AFOSC  & 9/10/01   & 0.473     & \#8      & 2.1  & 6250-8050 & 190 \\
TNG+DOLORES & 8/6/02    & 0.275     & LR-R     & 1.5  & 4470-10360 & 300 \\
Keck II+ESI & 10/6/02   & 0.168-0.120    & 175 lines mm$^{-1}$ & 1.0 &  3900-11000 & 35 \\
Ekar+AFOSC  & 17/6/02   & 0.473     & \#4      & 2.1  & 3360-7740 & 470 \\
Keck II+ESI & 5/9/03    & 0.168-0.120     & 175 lines mm$^{-1}$ & 1.0  & 3900-11000 & 35 \\
\hline\hline
\end{tabular}
\end{table*}
\section{Spectroscopic observations and data analysis}  
A program of optical imaging and spectroscopy was undertaken to observe our sample.
We acquired B and R images and low-resolution spectra of all the objects (106)
identified in \citet{Aus00}.
\begin{figure*}
   \centering
   \includegraphics[angle=-90,width=10cm]{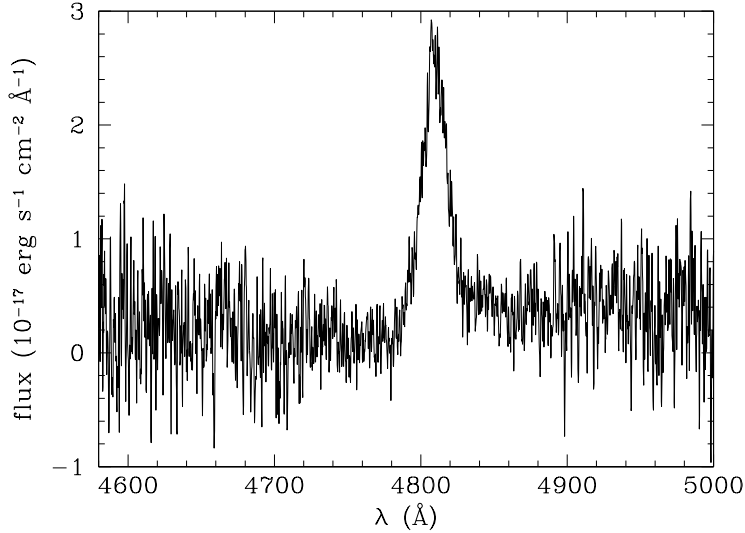}
\caption{Region of the spectrum of 3-40B source with the Ly$\alpha$ line visible.}
\label{Lyman}
\end{figure*}
Our targets were observed in different runs  in the years 2000 -- 2003 with telescopes and instruments whose set up is given in Table \ref{log_obs} (see also Table C.1 in \citetalias{DV05}).
Sixty-five IRAS fields were observed and spectra of 81 ISOCAM sources taken. In general the slit was oriented along the apparent major axis of the object. In those fields where a double component is visible, as discovered in our previous works (\citet{Aus00}, and \citetalias{DV05}), the slit was oriented  to take the spectra of both galaxies simultaneously.
We discovered that ten of our targets (3-04A, 3-10A, 3-44A, 3-49A, 3-57A, 3-65A, 3-78C, 3-83A, 3-85A, 3-89A) were physical pairs of galaxies with the same redshift estimate, and four objects (3-17A, 3-20B, 3-72A, 3-86A) were stars, so the total number of galaxies observed is 87, and out of them 85 have spectra with a good S/N. In this sample three galaxies (3-13A, 3-57A2, 3-68A) have an absorption line spectra, and six more objects show only a  very weak (S/N$\sim$2-3) H$\alpha$ emission line. For these galaxies we can measure the redshift, but the line is too faint to measure the flux. Finally one object, 3-40B, is a very peculiar case: its Keck spectrum, although very noisy, shows only one emission line that we identified as the Ly-$\alpha$ at $z=2.954$ (see Fig.\ref{Lyman}). For this ISOCAM  source, we only have upper limits both at $60\,\mu$m and at $15\,\mu$m \citep{Maetal01}. Thus the final sample that we analyze is composed of 75 galaxies. 

During each observing night some spectrophotometric and radial velocity
standard stars were observed with the same configuration grism/slit of the 
target objects. All the data were reduced using standard IRAF reduction packages\footnote{IRAF is distributed by the National Optical Astronomy Observatory, which is operated by the Association of Universities for Research in Astronomy (AURA) under cooperative agreement with the National Science Foundation.} and a detailed description of the reduction is given in \citetalias{DV05}. Spectra were classified according to the presence or absence of various emission lines. 
Observed emission lines included [O\,II]\,3727, H$\beta$, [O\,III]\,4959-5007,  H$\alpha$, [N\,II]\,6548-6583, and [S\,II]\,6717-6731. The measured redshifts are reported in \citetalias{DV05}.

In addition to the measure of the lines intensity, the high resolution of the ESI spectra allowed us to measure the rotation curves for almost all the galaxies observed with Keck~II ($31$ galaxies).

\subsection{Starlight subtraction} 
The optical spectra are contaminated and often dominated by the absorption lines of the stellar
component, which affect the strengths of most emission lines of interest. The magnitude of 
this effect depends on the equivalent widths of the emission and  absorption lines, 
and it is generally large in the nucleus of galaxies. For accurate measurements 
of their fluxes, we must remove the starlight contribution from our spectra. 
Generally, there are two strategies for removing starlight and thus obtaining a continuum  
subtracted, pure emission line spectrum: i) an off-nuclear spectrum (without emission lines) is subtracted from the spectrum of the nucleus, or ii) a template spectrum, free of emission lines, is properly scaled to and subtracted from the spectrum of interest.   

We used the second method, i.e., the technique of \citet{FH84} and \citet{hfs93}.
 For this purpose, we selected a sample of 57 pure absorption line galaxies from the 
data of the third data release of the Sloan Digital Sky Survey (SDSS; \citet{abz05}), 
whose  spectra have very similar characteristics to ours.  In particular, to account for
the underlying stellar continuum due to the bulge component,  template galaxies with different properties such as internal reddening and line strength Lick indices (Mg1, Mg2, Mgb, NaD), have been selected. The line strength indices are in the range typical of an old stellar population, and they include all the values found by \citet{Ann07}. 
The internal absorption ranges from zero to $A_v\sim$10 mag as derived from the H$\alpha$/H$\beta$ ratio in our sample.

We proceeded with the following steps: i) correct our own spectra and template 
spectra for Galactic reddening using the extinction  values of \citet{sfd98}, 
ii) de-redshift the spectra to zero, iii) change the resolution of the templates 
spectra to the resolution of the NEPR ones, iv) subtract the template 
spectrum from the object spectrum. For each NEPR spectrum, we applied this procedure 
using all the templates. The best subtraction was chosen using the $\chi^2$ minimization.
This method works successfully with the low-resolution spectra of TNG and Ekar telescopes, but it did not work properly with the spectra taken at Keck~II. 
This is due to the large difference between the resolution of the
Keck~II spectra and that of the template spectra. For this reason
we did not subtract the starlight continuum to the Keck~II spectra. This implies that fluxes of such emission lines cloud be underestimated.

\begin{figure*}
\centering
\includegraphics[width=12cm]{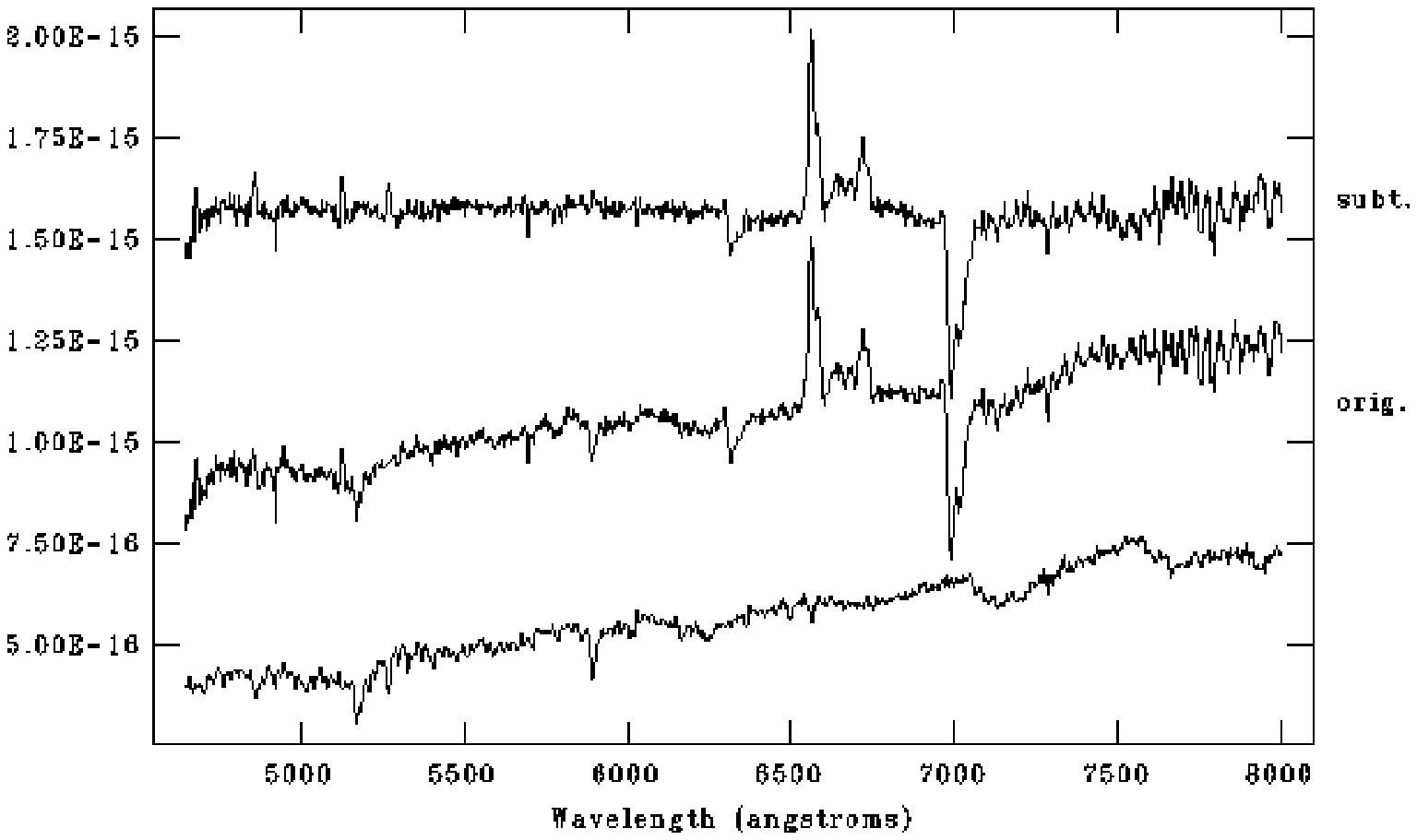}
\includegraphics[width=12cm]{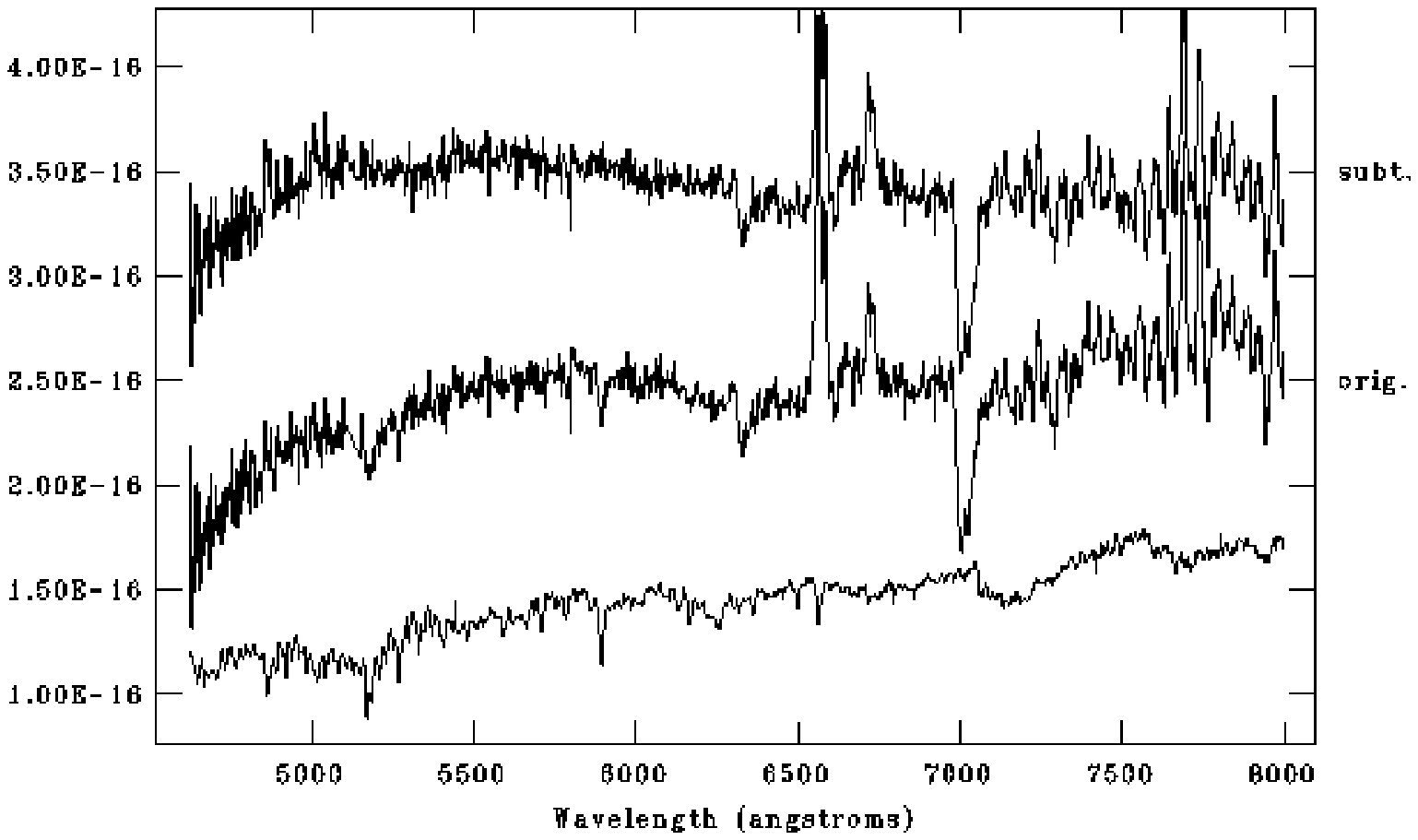}
\caption{Top panel: 3-38B, bottom panel: 3-77A. In each panel the bottom plot represents the best fitting template model used to match the stellar component. The middle spectrum is the observed one and the difference at the top. The spectra are scaled by an arbitrary constant.}
\label{conf}
\end{figure*}

In conclusion, for the Ekar and TNG spectra we used the fluxes 
corrected by the template subtraction. For spectra taken with the
Keck~II telescope, we used the raw fluxes, corrected only for Galactic
extinction. For the Ekar and TNG spectra, we continued our analysis quantifying
the error due to the choice of different templates. With this aim for each NEPR spectrum,  we compared the flux of H$\alpha$ emission line measured using the different template-subtracted spectra: we found differences from 2\% up to 10\%  in the case of high S/N and low S/N respectively.  Figure \ref{conf}  shows two examples of template-subtraction for high and low S/N spectra.

\subsection{Emission line flux measurement} 

In our observations the slit of the spectrograph was oriented along the apparent major axis of our targets or  with a position angle that allows having two galaxies along the slit. For this reason we have to take  differential atmospheric refraction (DAR) effects on our measured fluxes into account. To minimize these effects, our observations were made as close as possible to the zenith. In general our targets were observed with air mass $\sim$1.4,  and only few cases have higher values. Using the recipes in \cite{Fil82}, we derived the angular deviations to 
correct our fluxes for DAR, taking the air mass and the difference in parallattic angle into account. For the TNG and Ekar observations, we correct the data, were the angular deviation is inside our slit. For eight out of the galaxies observed at Keck~II, the conspiracy of both the small slit width (1") and high air mass ($\ge$1.8) give an angular deviation that is outside our slit and make DAR correction impossible. In Tables \ref{tab_em} and \ref{tab_em1} we mark these galaxies with an asterisk.
Moreover, the fluxes were corrected for reddening following the recipes in \citet{rosa02},  adopting the standard Milky Way value for R=3.1, and the intrinsic flux ratio of HII region like objects, F(H$\alpha$)/F(H$\beta$), 2.86.
The internal absorption ranges from $A_v \sim$1 mag to $A_v\sim$14 mag as derived from the H$\alpha$/$H\beta$ ratio in our sample. Its average value  is $A_v =6.7\pm3.0$\,mag with four highly attenuated objects. This is not surprising since our sample is composed of dusty galaxies whose far-IR properties were analyzed in \citet{Maz07}.   In Table \ref{tab_em} we give the $A_v$ values derived, when available.\\
This correction, as expected, does not affect the spectral classification.

We measured all the main parameters of the emission lines, i.e. 
fluxes with their errors (see Tables \ref{tab_em}, \ref{tab_em1}), fitting the emission lines with Gaussian functions. The  error estimates of the line fitting are computed by error propagation assuming independent pixel sigmas and no  errors in the background. In general in our spectra, the emission lines are also extended along the spatial axis. We decided to perform a measure adding together all the lines in a physical region of  3\,Kpc derived using the luminosity distance computed, for each galaxy, according to our cosmological model.  This is because, for our more distant objects  ($\sim$30\% of our sample) the 3\,Kpc aperture coincides with the whole optical extension of the galaxy. Thus, with this choice, all the sample is consistently compared.

\begin{figure*}
   \centering
   \includegraphics[width=8cm]{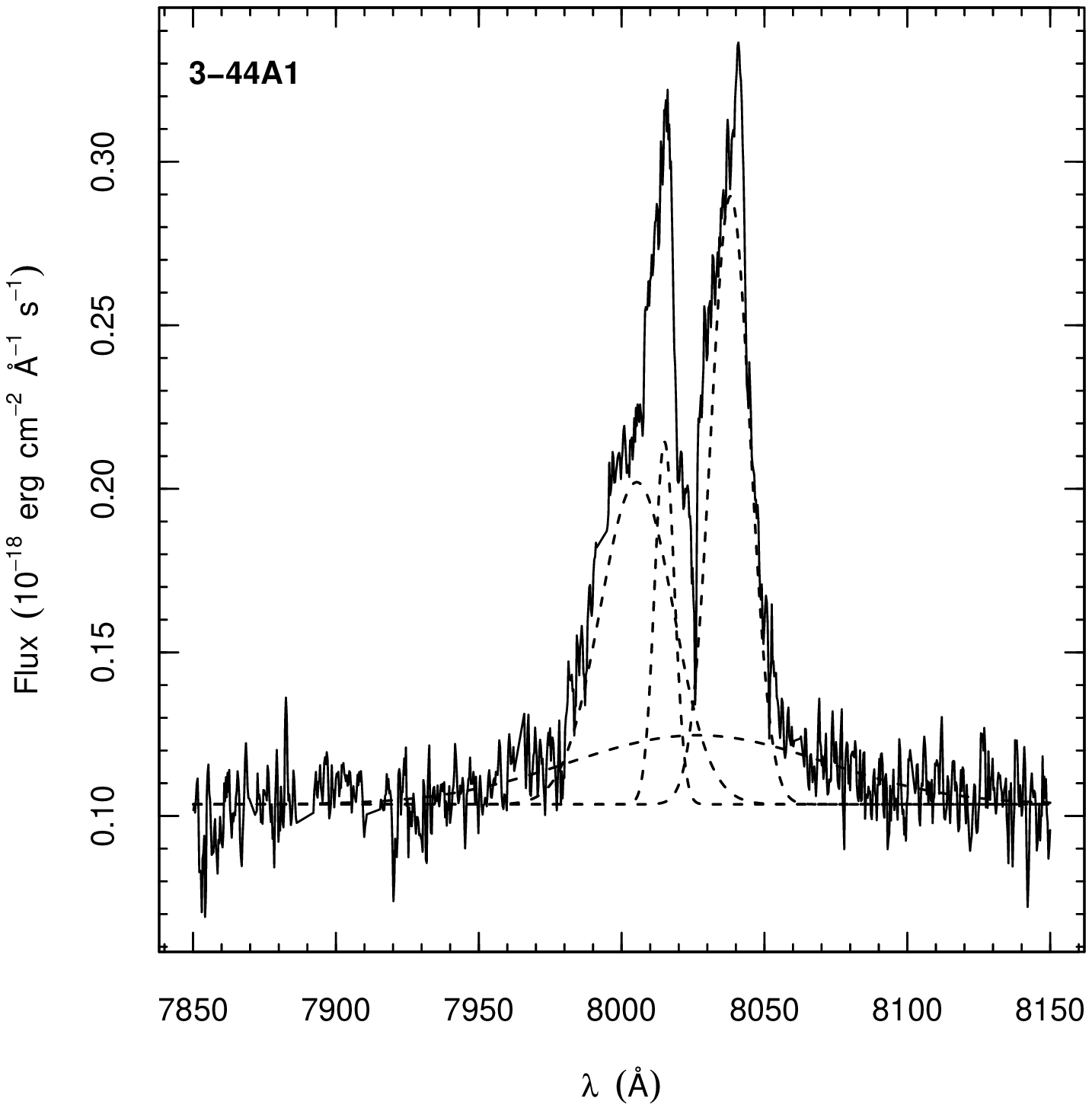}
   \includegraphics[width=8cm]{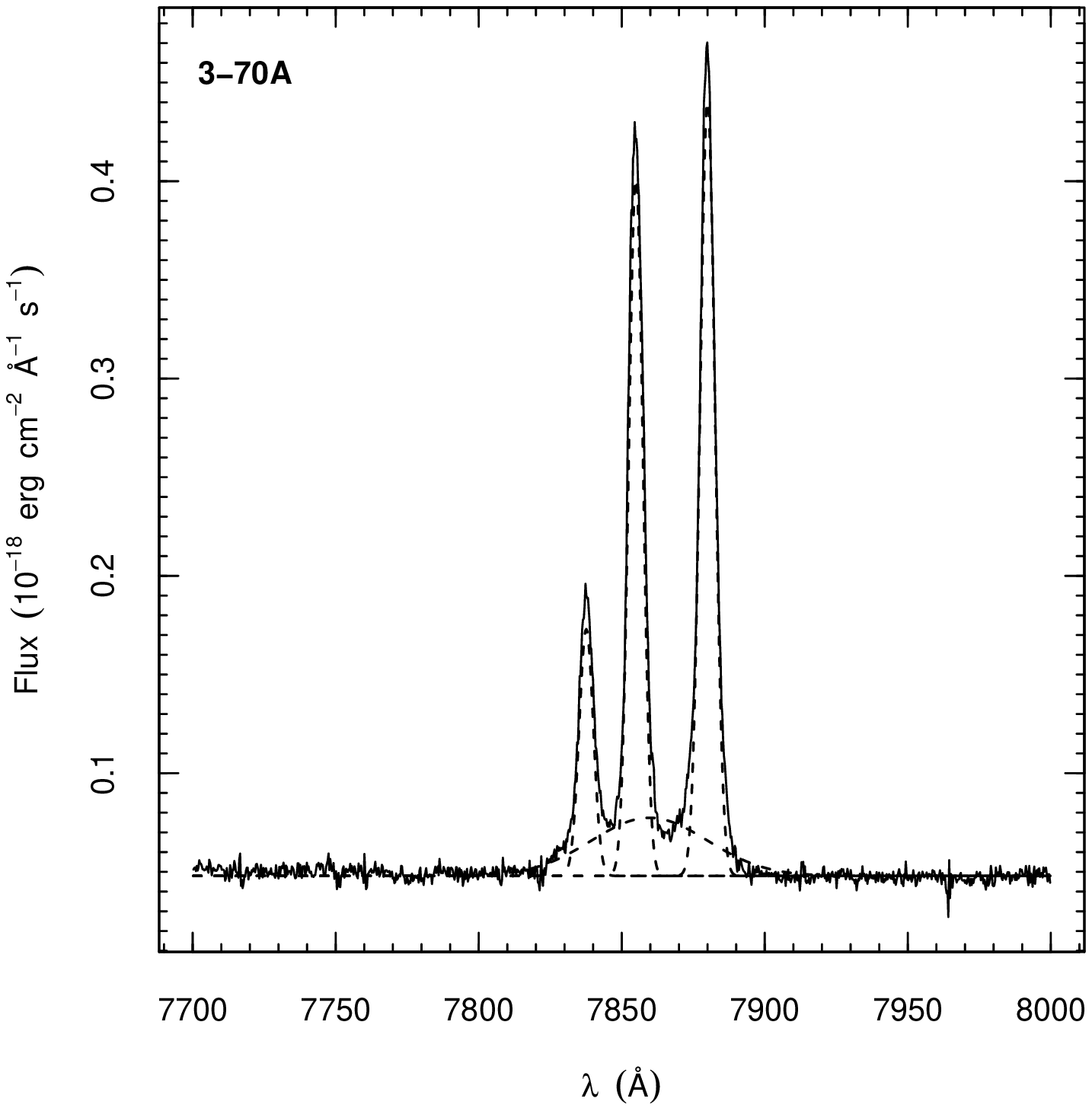}
   \includegraphics[width=8cm]{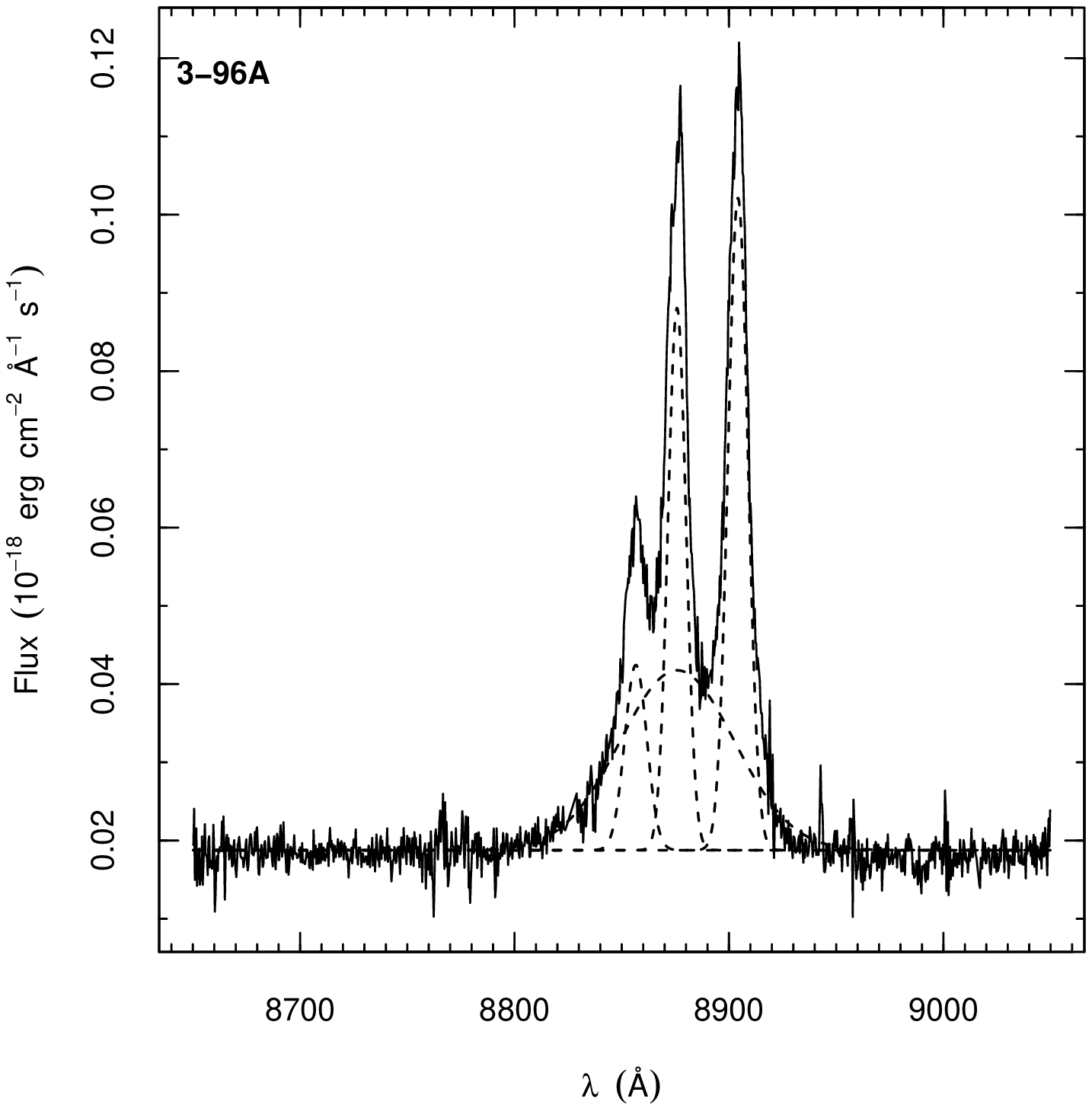}
\caption{ The H$\alpha$ region for the three AGN (upper left 3-44A1, upper right 3-70A, and bottom 3-96A). The dashed lines show our fits of the broad and narrow components.}
\label{AGN_spec}
\end{figure*}

In this {\it inner} region, the emission lines can be straightforwardly fitted by a single Gaussian function for almost all our sample. However there are some peculiar cases, as explained below:

\begin{itemize}

\item The emission lines of the spectra of 3-27A and 3-78C2 galaxies showed
a double peak. The  peaks' separation is  5.4~\AA ~and 3~\AA ~respectively, clearly indicating two kinematical components, as discussed in Sect. 4.1. For the lines of these two objects, 
two Gaussian functions were fitted, and the sum of their area was taken as total flux.

\item The spectra of the galaxies 3-44A1, 3-70A, and 3-96A exhibit a broad component in the H${\alpha}$
emission line typical of Seyfert~I galaxies, which was fitted by adding a second Gaussian function to the fit.
The FWHM and the flux of their broad line components are reported in Table \ref{broad_line}. Figure \ref{AGN_spec} shows their spectra and our fits in the H$\alpha$ region. 
\end{itemize}
Note that [NII]$\lambda$6548 was not fitted in the high-resolution spectra of 3-16A,
3-65A2, and 3-83A1, because a strong OH emission line of the sky was superimposed on it; however,
no flux correction was needed for the H${\alpha}$ fluxes of these three objects, since the H${\alpha}$ line and [NII] doublet are completely resolved in the Keck~II spectra.

When possible (47 spectra), the accuracy of the fit of the H${\alpha}$ group
was checked by evaluating the strength ratio [NII]$\lambda$6583$/$[NII]$\lambda$6548 that is expected to be $\sim3$. The values we obtained are in the range $2.5\div3.5$. 

\section{Spectral properties}
\begin{figure*}
   \centering
   \includegraphics[width=10cm]{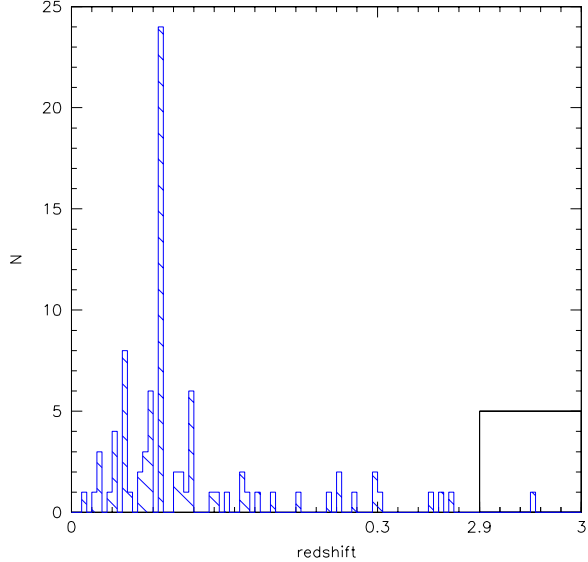}
\caption{Redshift distribution of our ISOCAM sources, in 65 IDS/ISOCAM fields. The insert in the lower right corner corresponds the
source 3-40B (see text).}
\label{redshift_d}
\end{figure*}
 Figure \ref{redshift_d} shows the redshift distribution of  all the 85 galaxies described in \citetalias{DV05}.
The NEPR supercluster (NEPSC) found by \citet{ash96}
\citep[see also][]{Burgetal92} dominates the $z$--distribution between 0.08 and 0.09.
Galaxy members of other expected clusters appear between 0.05 and
0.06, 0.07, and between 0.11 and 0.12. Sixty-one percent of our targets are galaxies with $z<0.1$, 25\% with 0.1$\le z \le $0.2, and 
14\%  galaxies with $z>0.2$. The mean redshift of the galaxies for which we can measure the rotation curves (see Section 4) is $\left<z_\mathrm{rot}\right> =  0.141$, and $\sim50\%$ of these have redshift $z \leq 0.1$.

Tables \ref{tab_em} and \ref{tab_em1} present the observed intensities, together with their errors, of the principal emission lines measured  inside the physical region of 3\,Kpc of our targets. Table \ref{tab_em} includes lines in the red spectral region, between 7000~\AA\, to 5000~\AA, and Table \ref{tab_em1} those in the blue range, between 5000~\AA\, and 3700~\AA.

\subsection{Spectral classification}
\begin{table}
\caption{Broad line data for 3-44A1, 3-70A, and 3-96A}
\label{broad_line}
\begin{tabular}{lcc}
\hline\hline
Galaxy  & FWHM & Flux \\
name    & km s$^{-1}$ & $10^{-18}$erg s$^{-1}$$cm^{-2}$$\AA^{-1}$\\
\hline
3-44A1  & 3700 & 3.712$\pm$0.896\\
3-70A   & 2260 & 1.416$\pm$0.083 \\
3-96A   & 2870 & 1.740$\pm$0.050 \\
\hline\hline
\end{tabular}
\end{table}

\begin{figure*}
\centering
\includegraphics[width=20cm]{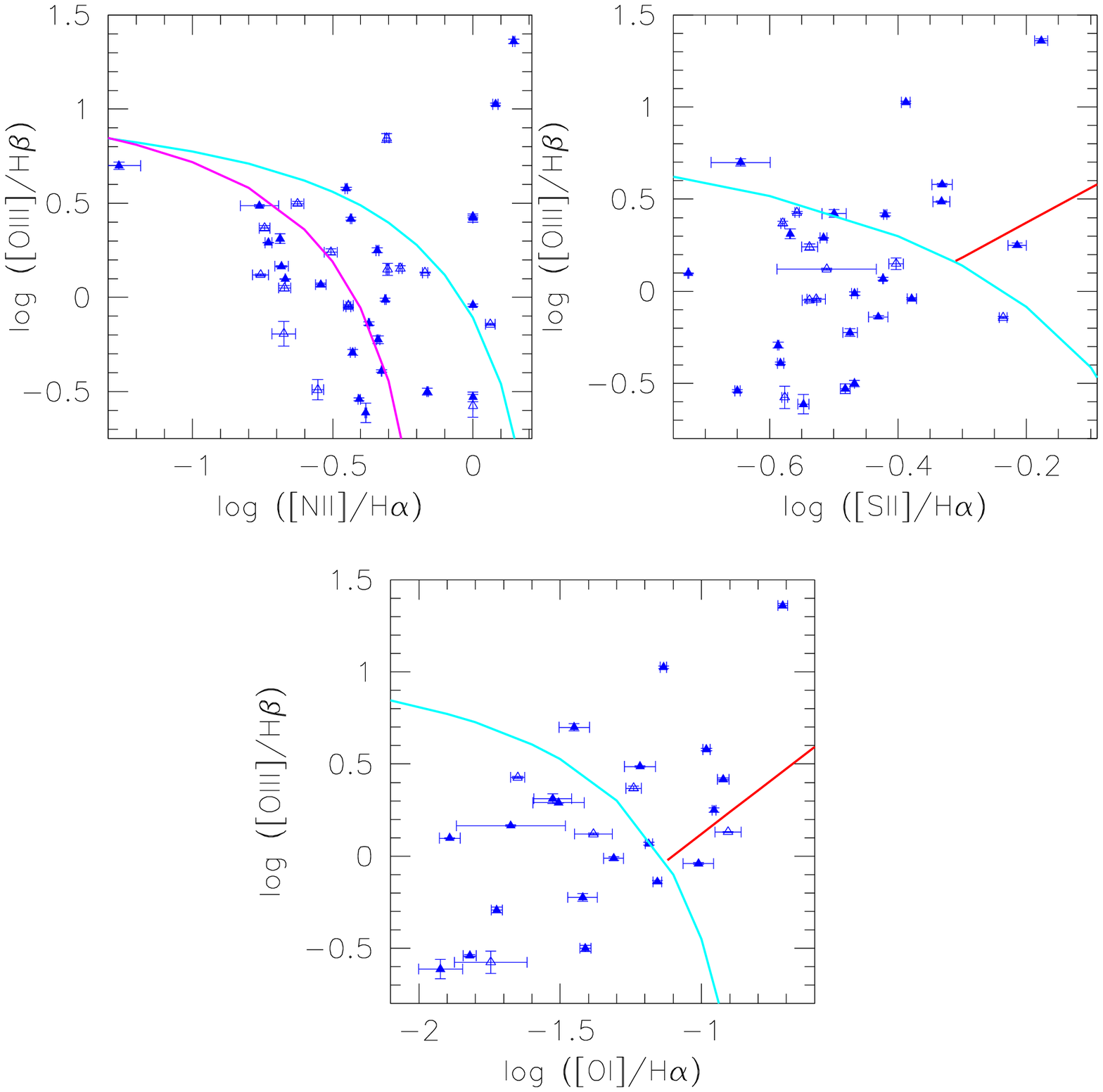}
\caption{ Emission line diagnostic diagrams. Galaxies belonging to the 60\,$\mu$m complete sample of \citet{Maetal01} are represented with filled triangles;  light (cyan) curves in all the three panels show the  extreme starburst definition of \citet{kew01};  bold curve (magenta) shows the pure SF limit of \citet{kau03}; and  red lines the LINER/AGN divisions from \citet{kew01} and \citet{kew06}. Upper panels: log [OIII] $\lambda$5007/H$\beta$ versus log [NII] $\lambda$6583/H$\alpha$ and log [OIII] $\lambda$5007/H$\beta$
versus log [SII] $\lambda$ 6716, 6731/H$\alpha$. Lower panel: log [OIII] $\lambda$5007/H$\beta$ versus log [OI] $\lambda$6300/H$\alpha$.}
\label{diagno}
\end{figure*}

The optical spectral classification  as  Seyfert (Sy), LINER (L), and star-forming (SF)  galaxies, is  based on diagnostic diagrams of \citet{vo87},  a revision of the pioneer work by \citet{bpt81}. They exploited four emission-line ratios of the most prominent bright emission lines: [OIII]$\lambda 5007/$H$\beta$, [NII]$\lambda 6584/$H$\alpha$, [OI]$\lambda 6300/$H$\alpha$  and [SII]$\lambda\lambda 6716,31/$H$\alpha$. These line  ratios take full advantage of the physical distinctions between the various types of objects and minimize the effects of reddening correction and calibration errors. Diagnostic diagrams have been used ever since as a standard to identify narrow--line active galaxy nuclei (AGN, or Sy).
\citet{kew01} (see also \citet{GK08} and references therein) build up a detailed starburst model with large ranges of metallicity and ionization parameter, finding upper limits to separate starburst from AGN on such diagrams. Star-forming (SF) galaxies fall onto the lower left-hand region of these plots, narrow-line Sy are located in the upper right, and Ls in the lower right-hand zone. Thus,
to separate the different types of galaxies (Sy, L, SF galaxies) we used the theoretical boundaries of \citet{kew01}.

Figure \ref{diagno} shows the position of our galaxies on these diagrams, including 
galaxies in the 60\,$\mu$m complete sample (S${60} \ge$ 80\,mJy) of \citet{Maetal01}. The data refer to the fluxes measured in the inner physical 3\,Kpc region of the galaxies. 
Three galaxies, 3-44A1, 3-70A, and 3-96A, show the H$\alpha$ with a broad line component (Fig. \ref{AGN_spec}), and we classified them as Sy~1. In Fig. \ref{diagno} they are residing in the AGN region and will be discussed further below.  Table \ref{broad_line} presents the FWHM and fluxes of their broad line H$\alpha$  emission, and Tables  \ref{tab_em} and \ref{tab_em1} the fluxes of their narrow line components used in the spectral classification.

In our sample there are 25 objects (33\%) with all the four  available diagnostic ratios (see Fig. \ref{diagno}). However, it is possible to address a spectral classification for other 17 galaxies using only two diagnostic diagrams, so we classify a total of 42 galaxies (56\% of the observed ones). 

After analyzing the diagnostic diagrams (Fig. \ref{diagno}) following \citet{kew06} and \citet{GK08}, all the galaxies below the pure SF line defined by \citet{kau03} are SF. We found that 16 out of 25  galaxies with  all four ratios  available, turn out to be SF galaxies.  

To classify all the remaining galaxies we used the theoretical one sigma boundaries between the regions occupied by LINERS (L) and AGN in these diagnostic plots, as in \citet{kew06}. 
We note that  two galaxies are certainly inside the region that defines the Ls, and seven more galaxies are in the region defined by the one sigma boundaries as Ls. The emission lines in the spectra of these seven galaxies  do not have a broad line component. This point further favors their classification as Ls. Following all these criteria, the
three AGN galaxies discussed above appear in the upper right-hand region of our diagnostic plots as expected. 


A tentative  spectroscopic classification of 3-84A was
made by \citet{ash92} obtaining an ambiguous result: they found that this galaxy has [NII]$\lambda$6583 emission indicative of a star-burst,
ambiguous [SII] emission, and [OI]$\lambda$6300 emission characteristic of an AGN. Following previous discussion we classify this galaxy as an L (see Table \ref{tab_sp_class}).
Finally, we found
\begin{itemize}
\item  30 SF (71\% and 40\% of the classified and observed galaxies
respectively),
\item  3 Sy~1 (7\% and 4\%)  

\item  9 L (21\% and 12\%) 
\end{itemize}

\begin{figure*}
\centering
\includegraphics[width=0.4\textwidth]{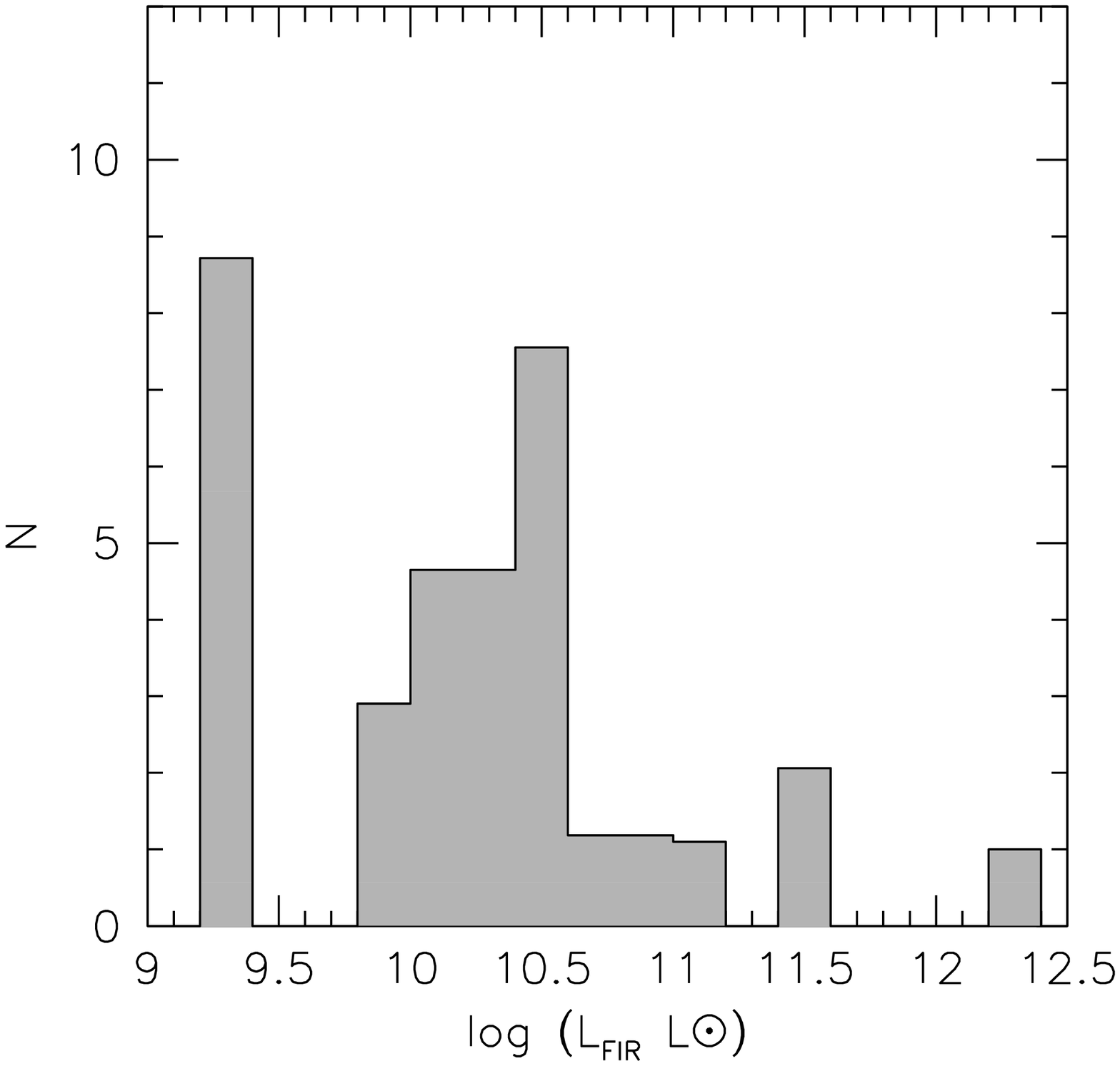}
\includegraphics[width=0.4\textwidth]{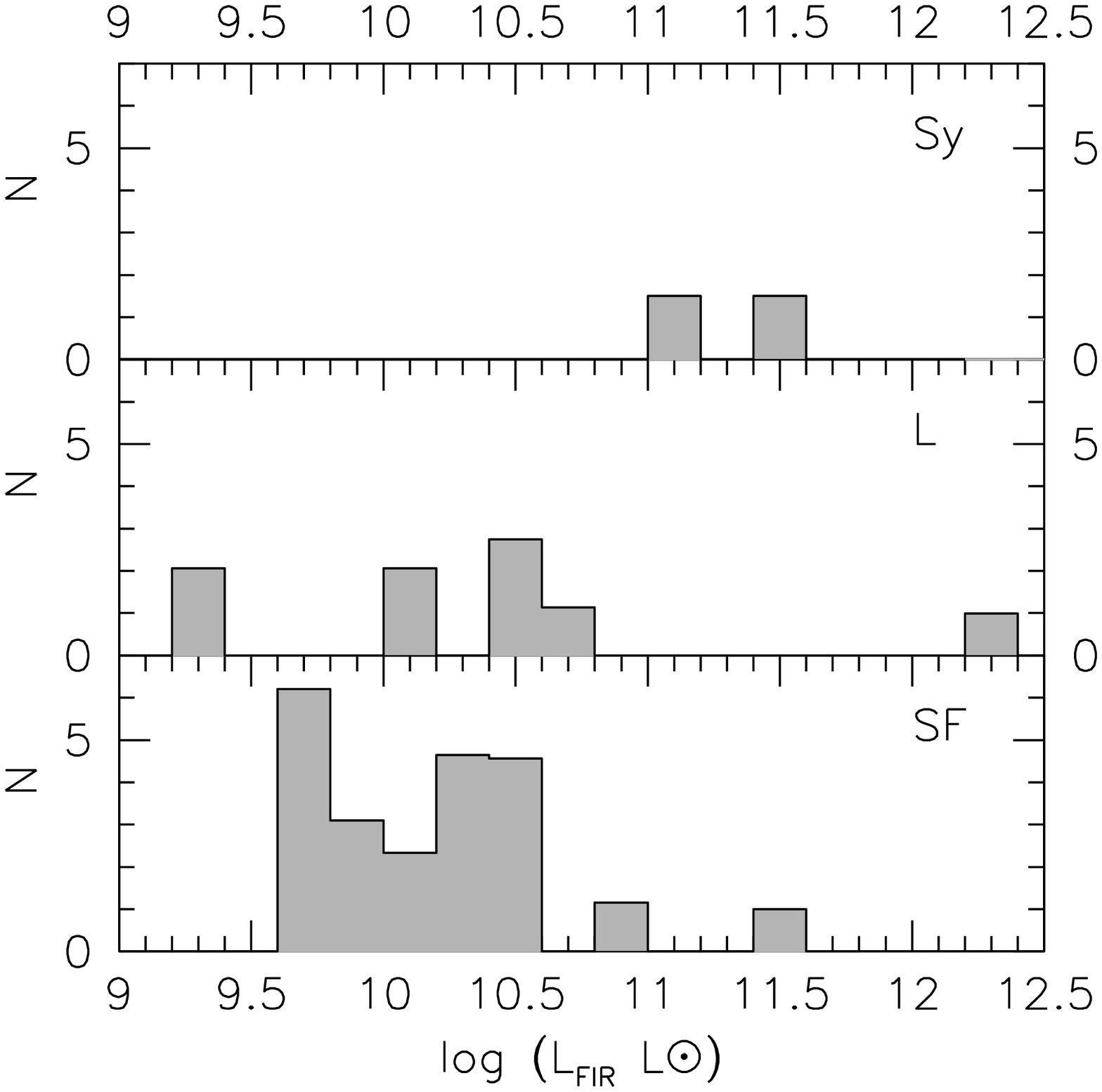}
\caption{{\sl Left}: The rest-frame FIR luminosity distribution of 35 ISO/IRAS sources included in this study. Far-IR data \citep{Maetal01} account for KM estimator (see text).
{\sl Right:} As in the left panel accounting for spectral classification here derived (Table \ref{tab_sp_class}).}
\label{class_FIR}       
\end{figure*}  

Table \ref{tab_sp_class} summarizes our spectral classification.
Figure~\ref{class_FIR} shows the distribution of the rest-frame FIR luminosity, L$_{FIR}$ from 42.5 to 122.5 $\mu$m, where

\begin{equation} 
L_\mathrm{FIR} = 4 \pi D^2 (FIR)
\end{equation}
\begin{equation}
FIR = 1.26 \times 10^{-14} (2.58 f_\mathrm{60} + f_\mathrm{100})\,W/m^2,
\end{equation}
 and f$_{60}$ and f$_{100}$ are in Jy \citep{he88}, of the 35 ISO sources corresponding to the 42 galaxies in Table \ref{tab_sp_class}. 
In particular, in Table  \ref{tab_sp_class}, 30 SF galaxies correspond to 26 ISO/IRAS sources, and three of these, 3-78C, 3-79C, and 3-92A, have no FIR fluxes \citep{Maetal01}, so there are 23 ISO/IRAS sources classified as SF types in the right-hand panel of Fig. ~\ref{class_FIR}. 
In this figure, as in the following ones, K-corrections were derived from evolutionary population synthesis models taking dust effects
into account  \citep{Maz95}, and luminosities were in units of solar bolometric luminosity, L=3.83$\times$10$^{33}$\,erg/s. Moreover, upper limits to flux densities were accounted for by exploiting the Kaplan-Meier (KM) estimator \citep{KM58}, as in \citetalias{Maz07}. Calculations were carried out using the ASURV v 1.2 package \citep{Isobe}, which implements methods presented in \citet{Feigelson1985} and in \citet{Isobe1986}.
The KM estimator is a nonparametric, maximum-likelihood-type estimator of the true distribution function (i.e., with all quantities properly measured, and no upper limits). The survivor function, giving the estimated proportion of objects with upper limits falling in each bin, does not produce, in general, integer numbers, but is normalized to the total number. This is why noninteger numbers of objects appear in the histograms of our figures.

The average value of the LFIR distribution, which accounts for 18 upper limits, is log(LFIR)=10.3 (Fig. \ref{class_FIR}, left panel), very close to the average value of the  FIR selected complete sample of \citet{Maz07}. Liner galaxies show the same average LFIR whereas, by accounting only for SF types, this slows down slightly, 9.9 (Fig. \ref{class_FIR}, right panel).

\section{The rotation curves}

\begin{figure*}
\centering
\includegraphics[width=0.40\textwidth]{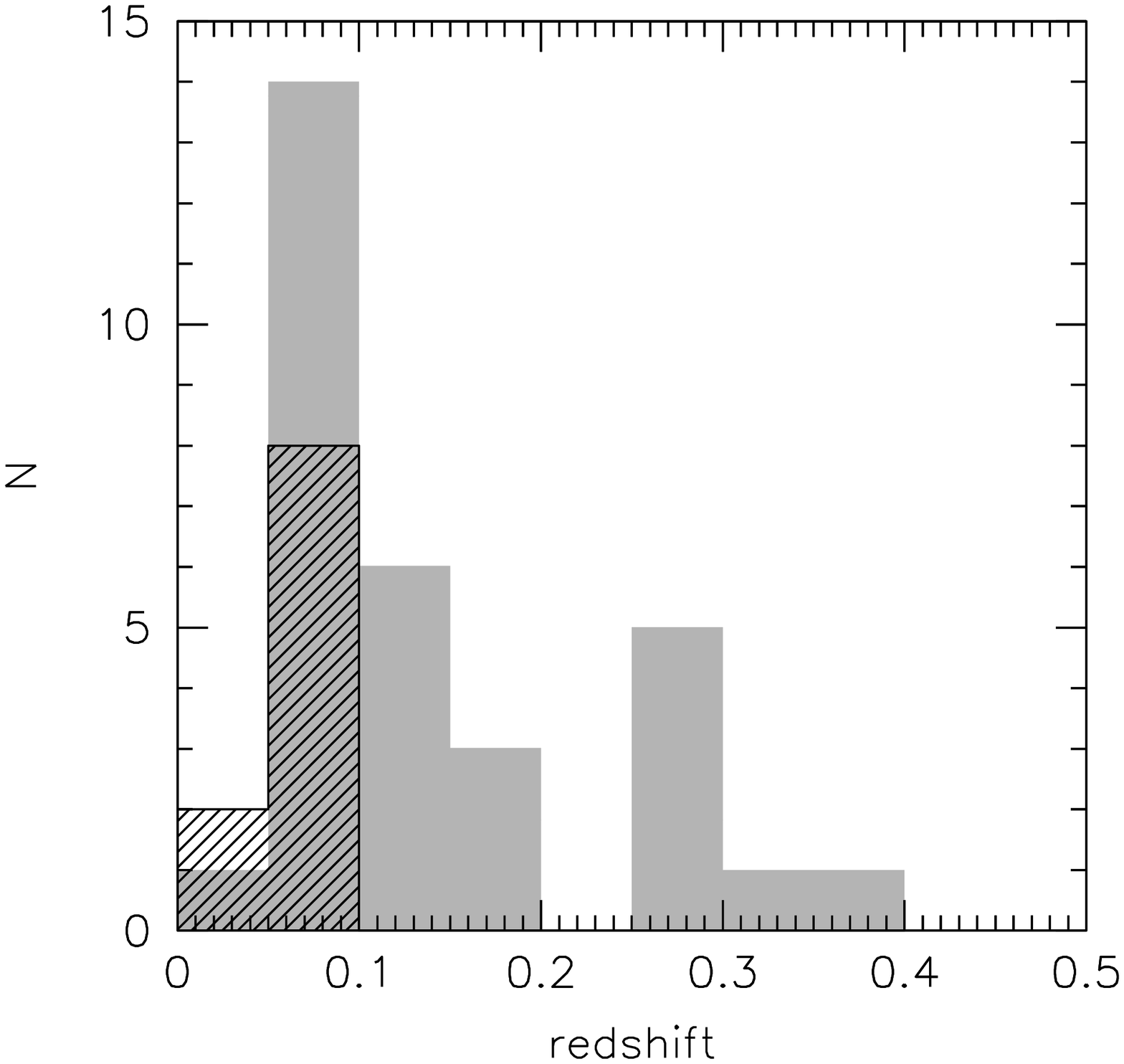}
\includegraphics[width=0.40\textwidth]{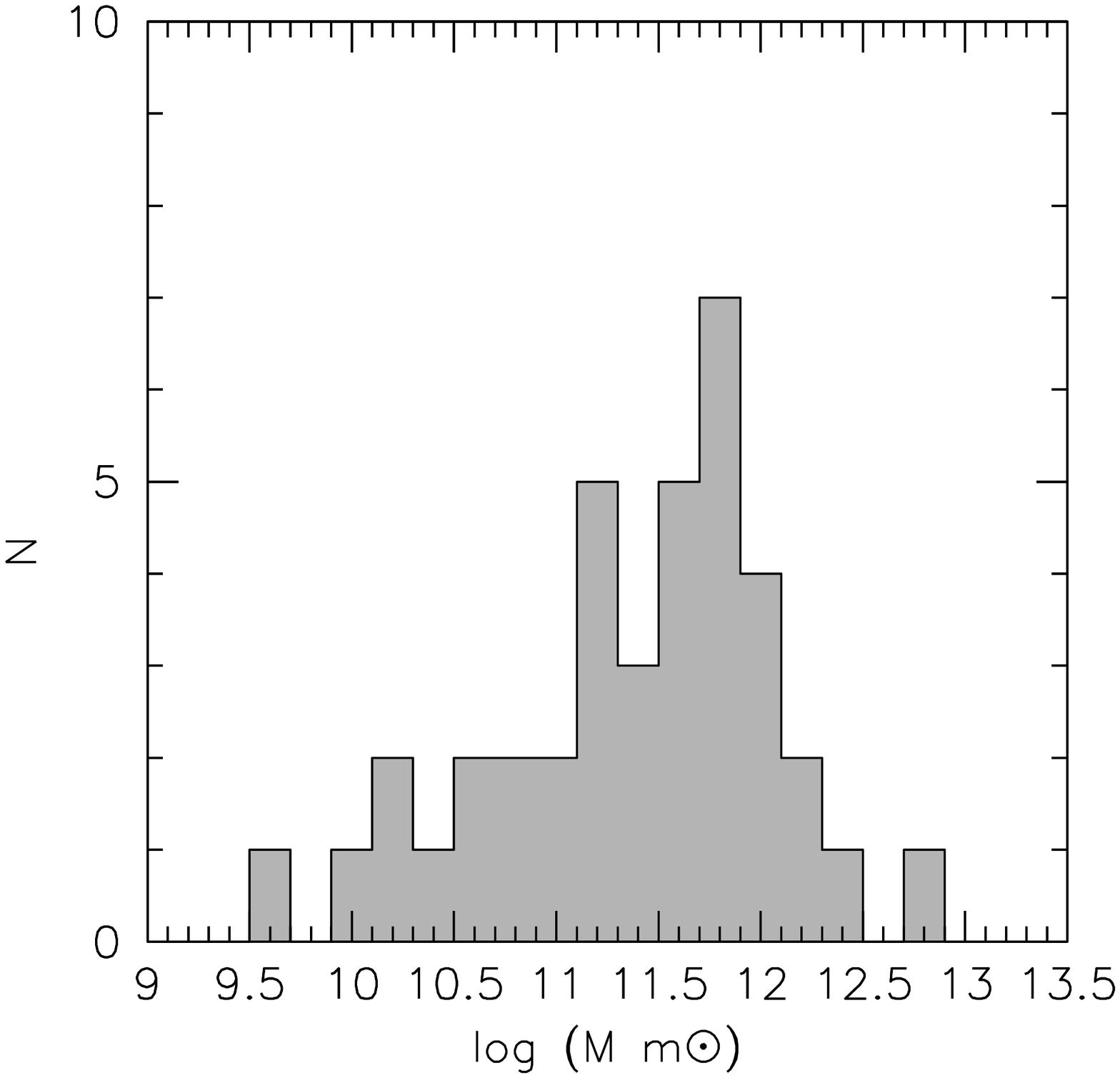}

\caption{\small Left: Redshift distribution of  41 galaxies in Table \ref{tab_rotcur}:  gray histogram is for 31 galaxies with rotation curves, dashed histogram  shows the remaining galaxies; the bin size is $\Delta z = 0.05$. Right:  Mass distribution of 39 galaxies in the same Table (see text); the bin size is 0.2.
}
\label{zrot}
\end{figure*}
\begin{figure*}
\centering
\includegraphics[width=0.40\textwidth]{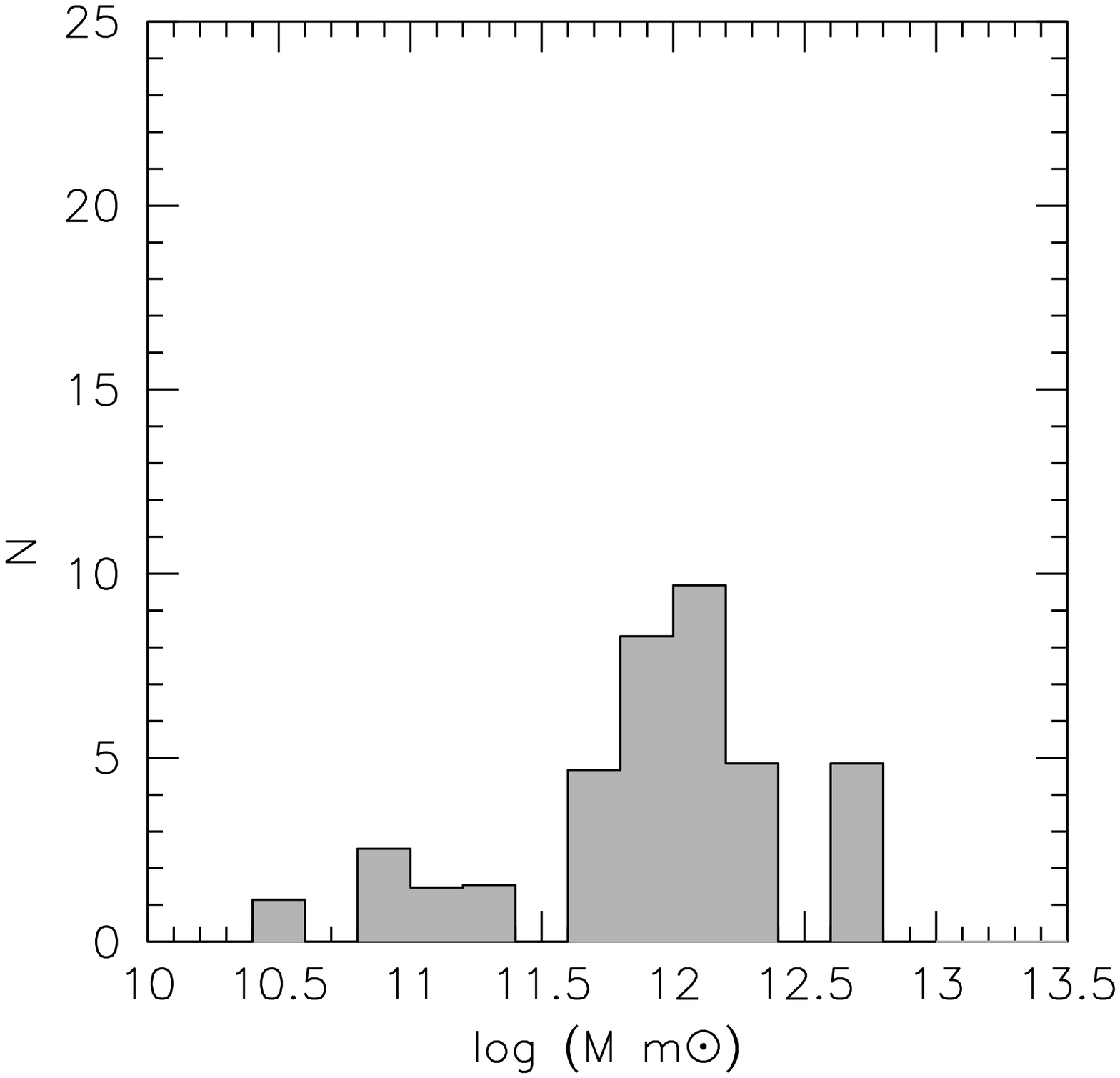}
\includegraphics[width=0.40\textwidth]{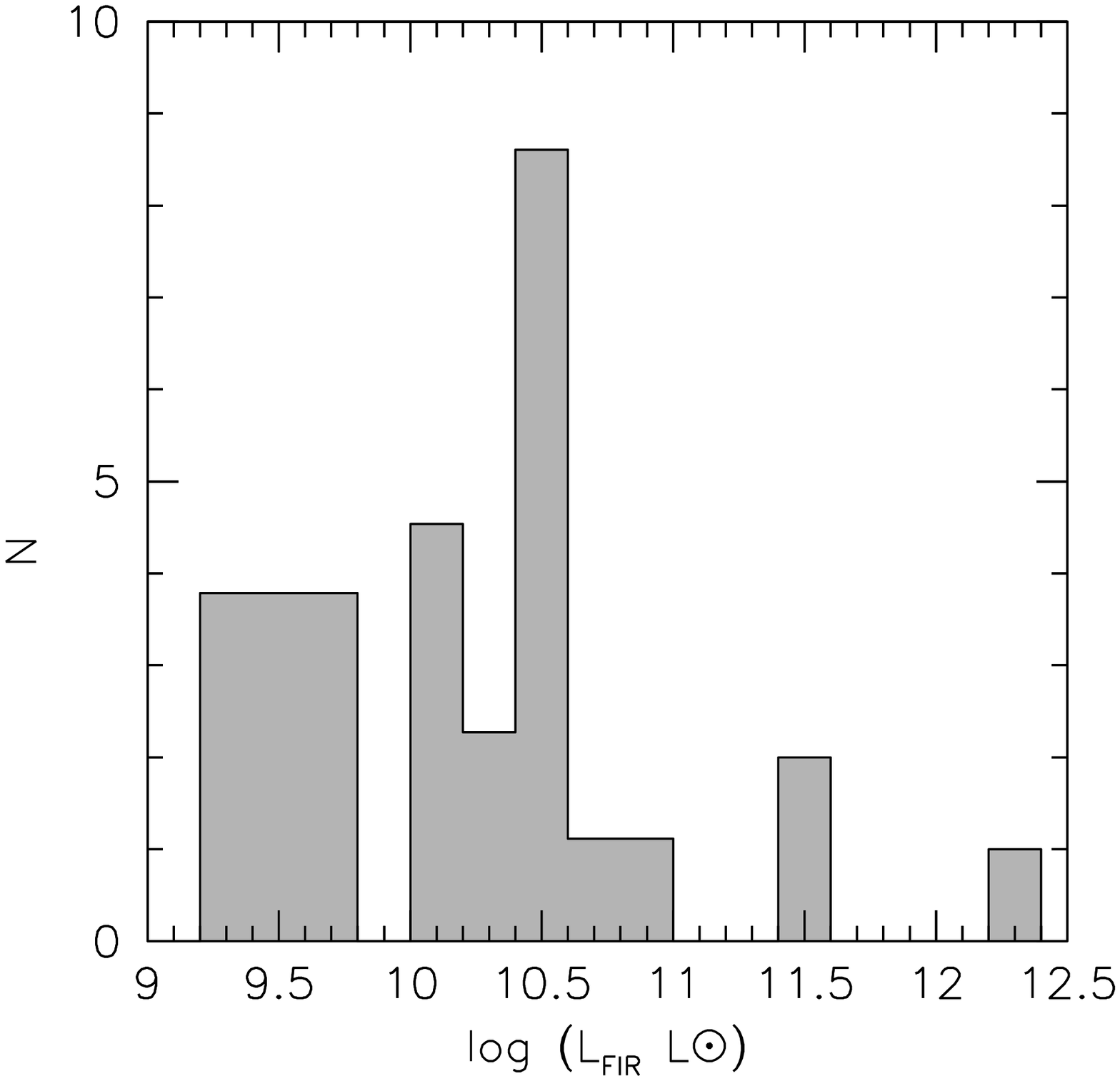}

\caption{\small Left: The same as in Fig. \ref{zrot} but acconting for KM estimator. Right: The rest-frame FIR luminosity distribution for 32 galaxies with mass estimates exploiting KM estimator to take upper limits to FIR fluxes into account \citep{Maetal01}.
}
\label{zrotbis}
\end{figure*}

The spectra obtained with the Keck~II telescope have both wavelength and  spatial 
resolution high enough to allow us to measure the galaxy rotation curves (RC). Only three galaxies do not show emission lines and for this reason we have such data for   31 out of the thirty-four galaxies observed with this instrument (41\%  of the whole sample of observed galaxies). 
To measure the RCs, a Gaussian function was fitted on the H$\alpha$ emission lines with S/N $\geq 2$. In  some cases more than one component was  detected, either because of the presence of two decoupled kinematic components  as in 3-26B and 3-27A or because of interacting regions of two galaxies (3-78C2).

\begin{figure*}
\centering
\includegraphics[angle=-90,width=0.9\textwidth]{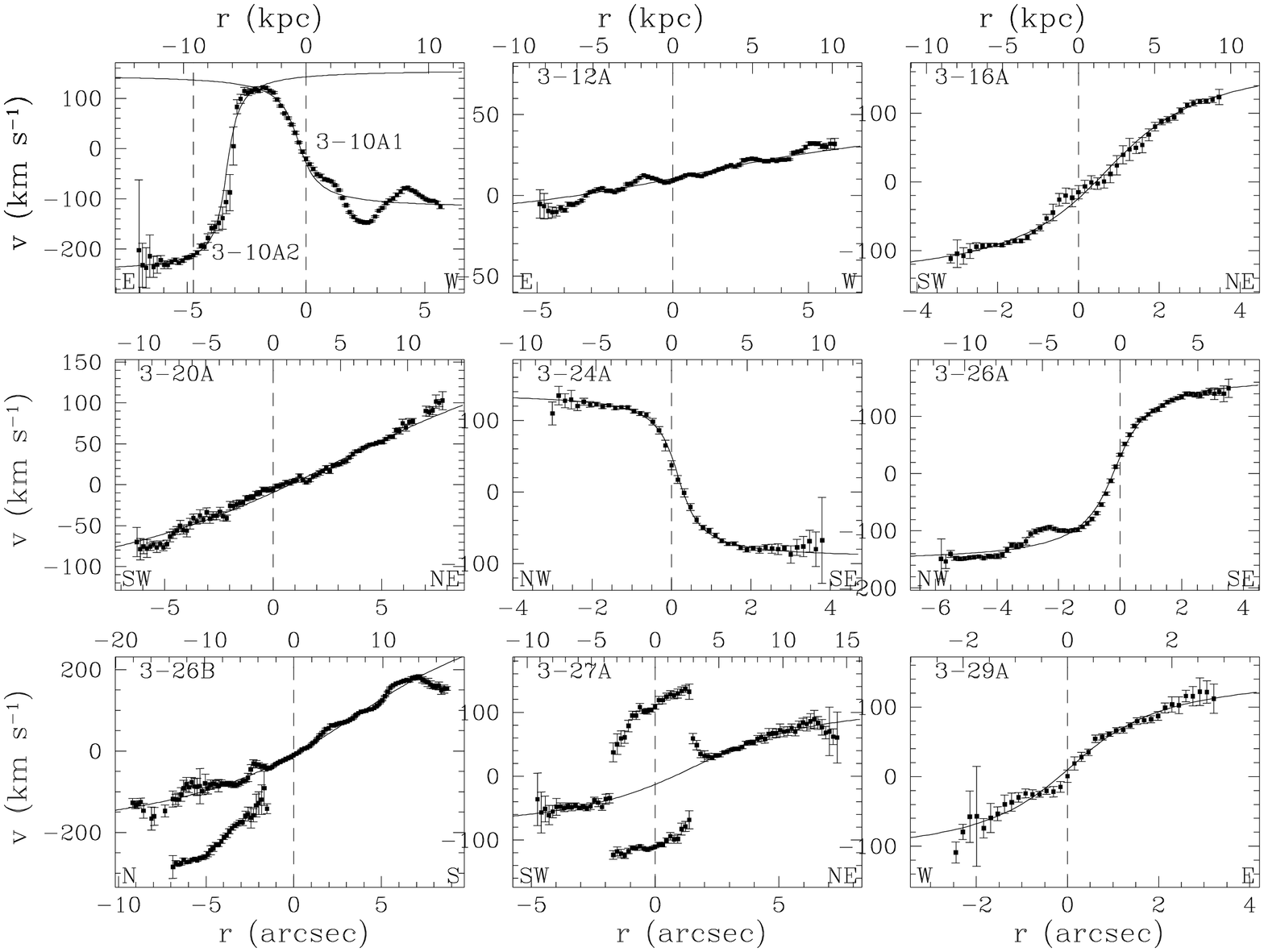}
\includegraphics[angle=-90,width=0.9\textwidth]{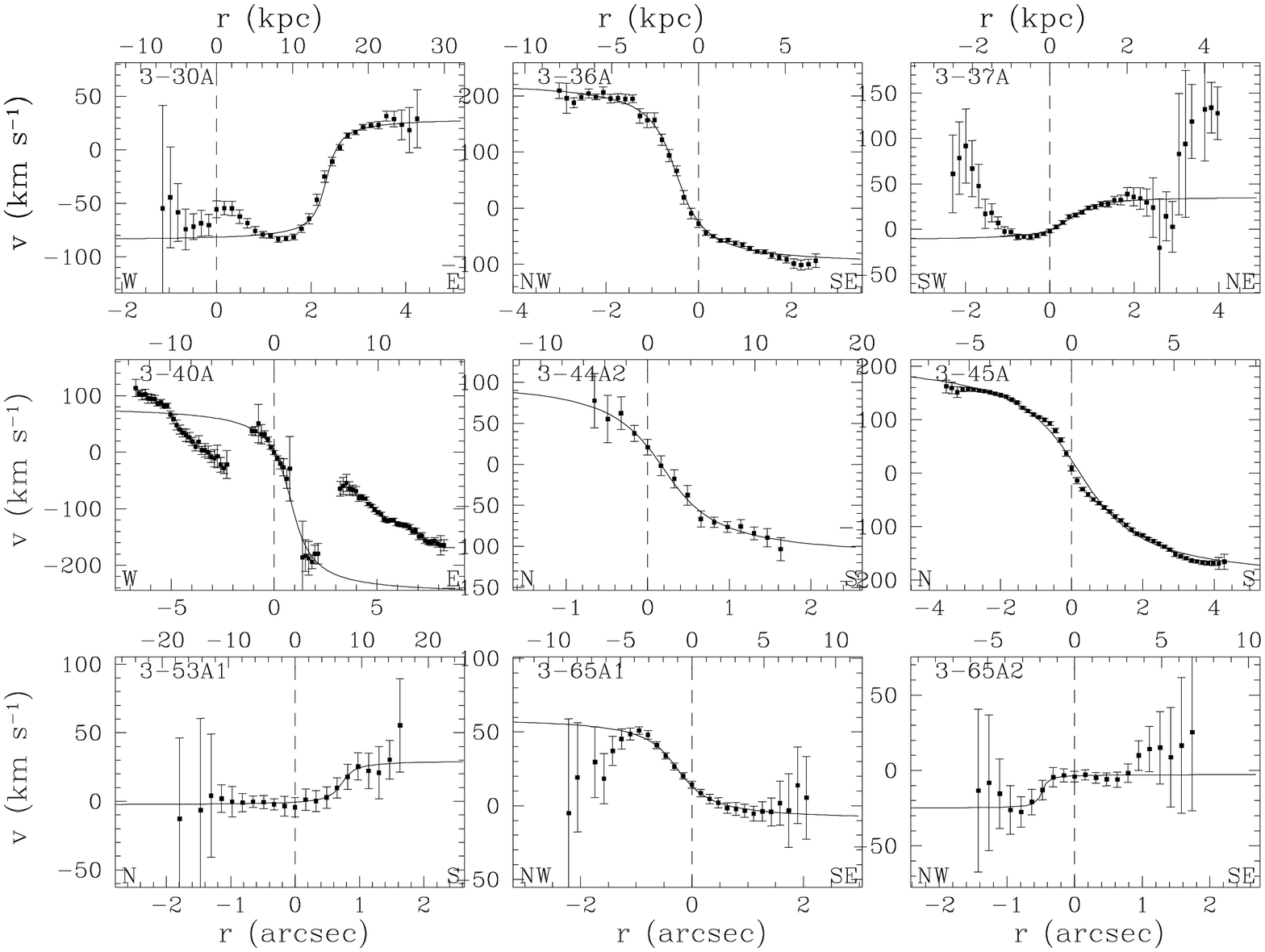}
\caption{The observed rotation curves for  the galaxies in our
       sample (at a mean $z\sim0.1$).  The solid line  shows the fit with  equation (\ref{eq:arctan}).}
\label{cr12}
\end{figure*}

\begin{figure*}
\centering
\includegraphics[angle=-90,width=0.9\textwidth]{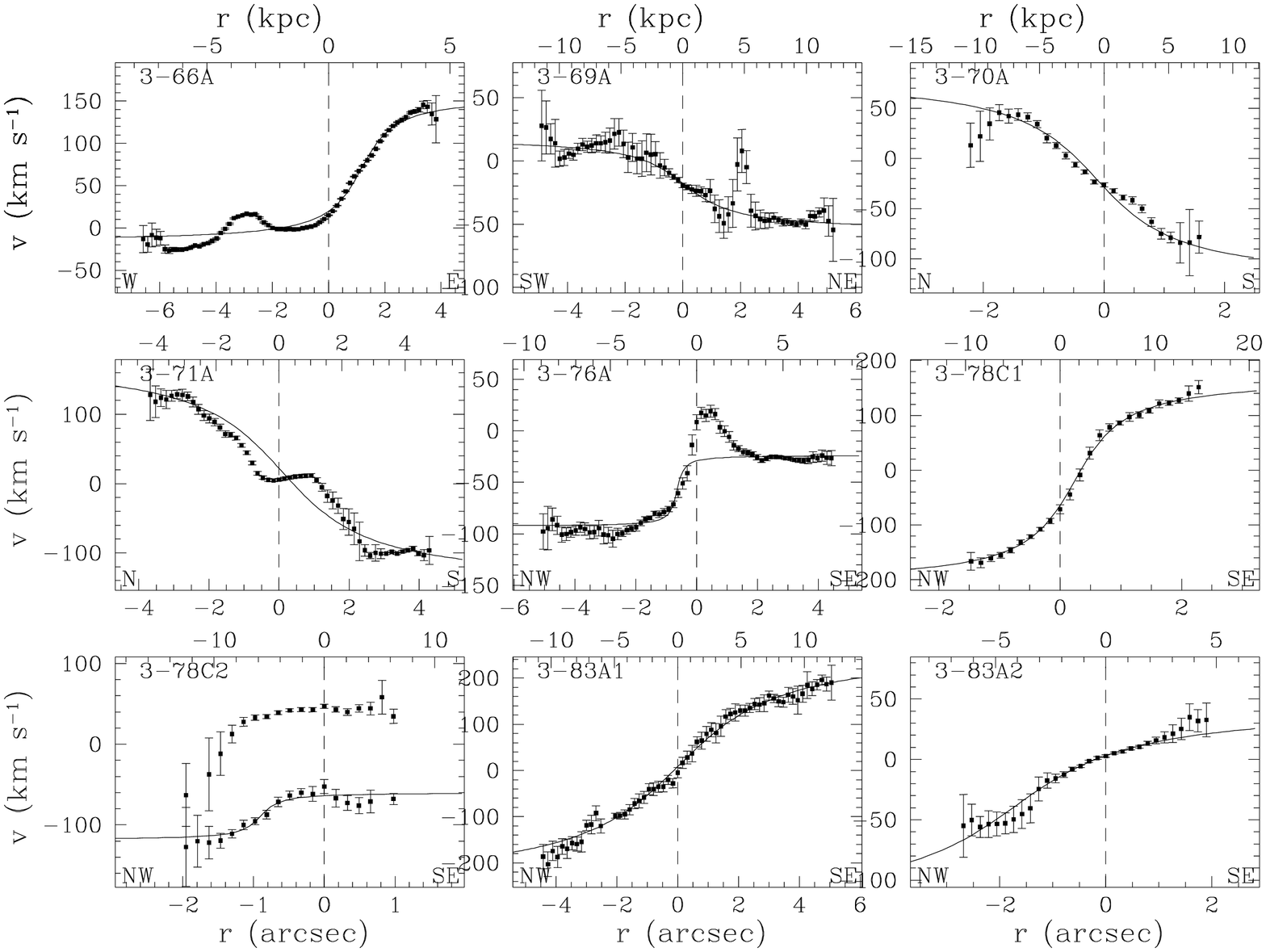}
\includegraphics[angle=-90,width=0.9\textwidth]{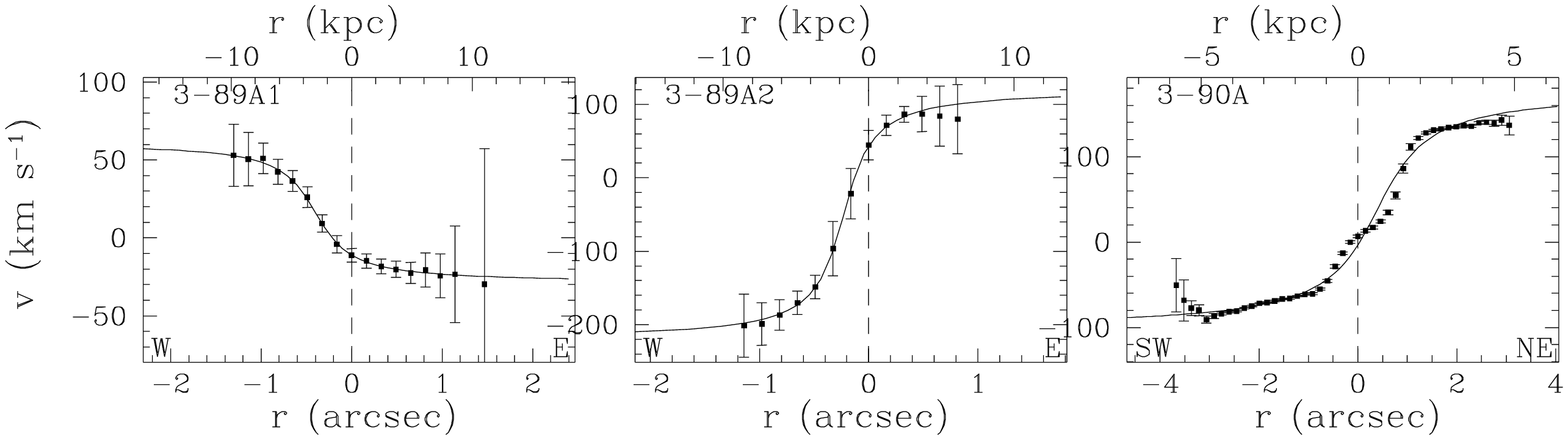}
\caption{The observed rotation curves for galaxies in our
       sample (at a mean $z\sim0.1$). The solid line  shows the fit with 
       equation (\ref{eq:arctan}).}
\label{cr34}       
\end{figure*}   

Since the main goal of our spectroscopic observations was to measure the
redshifts of the NEPR galaxies, the slit position of the spectrograph was not 
always placed along the galaxy major axis: for example, some objects were
observed simultaneously in the same exposure (see Section 2). The slit orientations therefore, have heterogeneous misalignment respect to the real major axes of the
galaxies: for nine out of them the misalignment is less than ten degrees, and
the slit was oriented along an oblique axis for the remaining 22 galaxies. This misalignment angle, $\delta$, has been measured for all the Keck~II spectra, and it is reported in Table \ref{tab_rotcur}. 
The redshift distribution of 31 galaxies for which we measured the 
RCs is shown in the left-hand panel of Fig.\ref{zrot}. The mean redshift of the
distribution is $\left<z_\mathrm{rot}\right> =  0.141$, the nearest galaxy is
3-29A with $z = 0.0408$, and the most distant one is 3-53A1 with $z=0.3723$.
Fifteen galaxies have a redshift $z \leq 0.1$. Two galaxies (3-44A2 and 3-53A1) have $z > 0.3$. 
The observed RCs have been modeled with the normalized {\tt arctan} rotation curve function:

\begin{equation} \label{eq:arctan}
v\left(r\right) = v_0 + \frac{2}{\pi} v_* \arctan\left(R\right)
\end{equation}

\noindent where $R=(r-r_0)/r_t$, and $v_0$ is the systemic velocity of the nucleus,
$r_0$ the spatial center of the galaxy, $v_*$  the asymptotic velocity,
and $r_t$  the transition  between the rising and flat parts of the
rotation curve \citep{Curt97}. In Figs.~\ref{cr12} and \ref{cr34},  the best fit  (Eq. \ref{eq:arctan}, continuous line) is shown.

To enlarge the sample of kinematical data in Table \ref{tab_rotcur}, we included the results from some 
low-resolution spectra obtained with the TNG telescope. 
These spectra do not allow the measure
of the RCs; nevertheless, since the spectra of the ten galaxies observed with this telescope (3-23A, 3-31A, 3-38A, 3-47A, 3-59A, 3-61A,  3-81A, 3-81B, 3-88A1, 3-93A) are spatially resolved, we succeeded in measuring at least the $\Delta v$,  i.e., the velocity difference between the core and the outer regions. 
The redshift distribution of these galaxies
is shown in the left-hand panel of Fig. \ref{zrot}. The mean redshift of the
distribution is $\left<z_\mathrm{rot}\right> =  0.063$, the nearest galaxy is
3-31A with $z = 0.0267$, and the most distant is 3-38A with $z=0.0883$.
Usually, we extracted 
regions around the core of $1.4''$ with the exception of 3-59A ($1''$), 3-88A1
($1''$), and 3-93A ($2.24''$). 
Then, we extracted spectra of  two regions in the outer part of the 
galaxies. 
For each of these spectra, the position of the blending group of H$\alpha$ and [NII]
emission lines was measured. For four galaxies (3-23A, 3-31A, 3-59A, 3-61A),
these three emission lines were deblended, while in the
remaining  ones only the positions of H$\alpha$ and [NII] were measured. Since 
[NII] is much weaker than H$\alpha$, we considered its contribution negligible.
Finally, the velocity difference, $\Delta v$, between the velocity of the
nucleus and in the outer region was calculated. 

 \subsection{Notes concerning peculiar RCs}
Most of the thirty-one   RCs in Figs.~\ref{cr12} and~\ref{cr34}
are very regular, which is typical of disk galaxies; however, some show
peculiar motions,  very disturbed rotation patterns, and in few cases,  kinematic decoupled regions that were excluded from the fit of RC.
In the following we summarize the puzzling features.

{\bf 3-10A}: this ISO/IRAS source is composed of two galaxies, A1 and A2, at both the same redshift \citep{DV05}.  Their disturbed RCs show these  are interacting/merging galaxies.

{\bf 3-12A}: its RC  extends beyond 8\,Kpc, but only the region of rigid rotation is visible. No imaging is available to confirm this point.

{\bf 3-20}: also in this case, the  RC  extends more than 9\,Kpc, showing
only the rigid rotation zone; there is no imaging available for this source either.

{\bf 3-26B}: this is a edge-on spiral galaxy. Its RC shows a double component in the northern region, towards the companion galaxy 3-26A. In the GALEX archival images,  a bridge connecting the two galaxies appears in the same region where the RC curve shows the double component; the plateau of its RC is hinted at because the slit of the spectrograph, placed along the major axis, is smaller than the apparent dimension of such a galaxy. 

{\bf 3-27A}: two inner velocity components separated by more than 200\,km/s arise in the RC of this galaxy. They  correspond to the opposite sides of a fast rotating ring decoupled from the disk motion.

{\bf 3-30A}: the west side of its RC, with the photometric center 2.2arcsec  far from the dynamical center, shows perturbed behavior. This pattern may be a signature of  an outburst or of some type of interaction; however, no imaging is available to explain the peculiarity of such an RC.

{\bf 3-37A}: this galaxy corresponds to PGC3086419 (HYPERLEDA catalog), but also in this case no imaging deep enough is available to solve the puzzle of its RC.

{\bf 3-40A}: even if this galaxy corresponds to PGC2699424 (HYPERLEDA) and to 2MASX J17562643+6723549, the resolution of available images cannot explain the very perturbed RC of this galaxy that could be the signature of a fast rotating inner bar.

{\bf 3-53A}: this is the highest redshift (0.37) galaxy belonging to the 60$\mu$m complete (S$_{60}>$80\,mJy) sample \citep{Maetal01}, and its RC only refers  to the slowly rotating  inner region. 

{\bf 3-65A}: this ISO/IRAS source is composed by a pair of galaxies, A1 and A2 \citep{DV05}; the north-west distortion of RC of A1 galaxy  cloud be due to the interaction with its companion galaxy, A2; however, our imaging do not confirm tidal distortions.

{\bf 3-66A}: there is an anomaly in the western region of its RC, which can be explained, looking at the DSS image (the only available so far), by an outburst visible in the north-west side of this galaxy.

{\bf 3-69A}: the anomaly arising in the north-east region of its RC  may be due to a companion galaxy. No imaging with enough resolution is available to elucidate this point.

{\bf 3-76A}: in the south-est region of its RC a perturbation arises that, considering the 2MASS image (2MASXJ18042694+6720481), may be due to a faint object near the galaxy.

{\bf 3-78C2}: there is a double component in the RC of this galaxy that may be due to interaction; however, its companion galaxy, 3-78C1, shows no signs of tidal interaction.

We derived the spectral classification of 28 out of the 31 galaxies with RCs. For three objects, namely 3-40, 3-45, and 3-76, no [OIII]$\lambda$5007 is detected. Twenty-three galaxies (77\%) are classified as SF,  four (13\%) as L, and one as Sy. We note that Ls and Sy  show an almost regular RC, while the SF galaxies show the distortions and peculiarities typical of interactions.
\subsection {Mass estimates}

We defined $v_{\mathrm opt}$
as the velocity measured at the maximum spatial extent in arcsec
($r_{\mathrm opt}$) of either the approaching or the receding side of the
galaxy spectra. No rotation curve shows the flat region of the curve clearly. 

Velocities are corrected for projection on the sky, cosmological stretch, and misalignment angle. Therefore, the corrected $v_c$
is there

\begin{equation} \label{eq:vcorr}
v_c = \frac{v}
{\left(1+z\right)\, \sin i\, \cos\delta} 
\end{equation}

\noindent where the inclination along the line of sight, $i$, was estimated
using the observed axis ratio derived from the CCD images when available (see \citetalias{DV05}), and the POSS images for the others; an intrinsic flattening, $q_0$, of 0.20 for all the galaxies is assumed. We use

\begin{equation}
\cos^2 i = \frac{q^2 - q_0^2}{1-q_0^2}
\end{equation}

\noindent where the adopted $i$ for each galaxy is given in Table~\ref{tab_rotcur}. 

The misalignment angle is a critical parameter for our measurements, because
several spectra were  observed far from the major axis. Following the
simulation of \citet{Gioetal97}, errors due to this correction are
negligible for position angle offsets less than $15^\circ$.
 For seventeen galaxies this condition is verified.
For  two,  3-65A2, and 3-70A,  this correction is instead not
applicable, because they were observed very close to the minor axis. For other two galaxies, 3-10A2,  and 3-81A the mass estimates is affected by this correction 
 (see Table \ref{tab_rotcur}).

Using the derived velocity $v_{\mathrm{c}}$,  we 
estimated  the mass  inside the last observed point of thirty-nine  galaxies, i.e.,
a lower limit to the mass of our galaxies, with the relation by \citet{Bosch2002}:
\begin{equation}
M_\mathrm{vir} = 2.54\cdot 10^{10} M_\odot \,
\left(\frac{r_d}{\mathrm{Kpc}} \right) \,
\left(\frac{v_\mathrm{c}}{100\, \mathrm{km s}^{-1}}\right)^2
\end{equation}
where $r_d$ is the virial radius. We used the radius $R_{opt}$ in Kpc as defined above as $r_d$.  These estimates are summarized in Table~\ref{tab_rotcur}.

The right-hand panel of Figure \ref{zrot} shows the mass distribution of galaxies in Table \ref{tab_rotcur}
and Fig. \ref{zrotbis}. The left-hand panel shows the same, accounting for KM estimator assuming  lower mass limits when the RCs are less extended than 8\,Kpc, and for all the $\Delta$v estimates.
The mass distribution spans more than three orders of
magnitudes. For  eleven galaxies the masses are derived in the  nuclear region, i.e., ($R_{opt}<$5\,Kpc).  The mean value of the distribution accounting for only RCs, i.e. 31 galaxies, is 3.66$\cdot 10^{11}M_{\odot}$, and 4.93$\cdot 10^{11}M_{\odot}$ including all the galaxies. 
 We point out that the high velocity, v$_{opt}$, measured for the 3-61A galaxy (see Table \ref{zrotbis}) may be due to dynamical perturbations induced by interaction  with a dwarf compact companion $\simeq$11.5 arcsec north of this barred galaxy,  as derived by combining the imaging and spectrum of this ISO/IRAS source. 
By removing this galaxy from the mass sample,  the last bin in Fig. \ref{zrot} (right panel) and in  Fig. \ref{zrotbis} (left panel) disappears, so that the mean value of the whole mass distribution is 3.68$\cdot 10^{11}M_{\odot}$, the same as derived from galaxies with RC measured.

We note that the 67\% of these galaxies belongs to the 60 $\mu$m selected, complete sample  defined by \citet{Maetal01}. The right-hand panel of Fig. \ref{zrotbis} shows the rest-frame FIR luminosity distribution of 32 ISO/IRAS sources corresponding to galaxies  in Table \ref{tab_rotcur}. Four galaxies  i.e., 3-10A2, 3-78C2, 3-83A2, and 3-83A2 are multiple optical counterparts of the same ISO/IRAS source \citep{DV05}; moreover, no FIR fluxes are available for 3-38A and 3-78C and 3-81B which result as confused sources in  \citet{Maetal01}. \citet{Maz07} show that the FIR luminosity of the IDS/ISOCAM sample extends over three  orders of magnitude (see their Fig. 9, left panel), as in Fig. \ref{zrotbis}, and with the same mean value, log(LFIR)=10.2. 
This is almost the same as for the Revised IRAS 60$\mu$m Bright Galaxy Sample \citep{San03} and for a normal spiral galaxy, like the Milky Way \citep{Metal92}. The ultraluminous infrared galaxy 3-53A emits the maximum FIR luminosity of the sample, nearly  100 times higher than the mean value.

\section{Conclusions}

The  spectral classification of 42 galaxies, with the emission line ratio diagnostic diagrams, shows that the NEPR sample is predominantly composed of SF, starburst galaxies (71\%), while the fraction of  LINERs (21\%) and AGNs (7\%)  is
smaller. Three new Sy 1 galaxies were identified,  3-44A1, 3-70A, and 3-96A.  The rest-frame FIR luminosity distribution of galaxies with spectral classification spans the same range as the FIR-selected complete sample analyzed by \citet{Maz07}, i.e. three orders of magnitude, with the same mean value, log($L_{FIR}$)=10.2. This emphasizes that such galaxies represent FIR properties of the whole sample well. Moreover, their optical properties are typical of the sample itself since 62\%  of these  belong to the 60\,$\mu$m selected complete sample of galaxies defined by \citet{Maetal01} (see \citet{DV05}).

Using the rotation curves and spatially resolved, low-resolution
spectra, we are able  to derive  dynamical parameters of 41 galaxies and mass estimates, inside the last point viewed,  of  39 galaxies in the sample.
We point out that the 67\% of them belong to the 60 $\mu$m complete sample  cited above. Moreover, also in this case, their rest-frame FIR luminosity distribution extends over the same range and has the same mean value as expected for the complete sample.
The mass distribution extends over three orders of magnitude with a mean value of $\left<M\right>=3.66\cdot 10^{11}M_{\odot}$,
slightly more than the Milky Way, where \citet{Wong2004} find a mass of 1.3$\cdot$ 10$^{11}$M$_{\odot}$ within 12\,Kpc  of the Galactic center.

\citetalias{DV05} concluded that two or more galaxies with very close redshifts may contribute to the ISOCAM/IRAS  flux in at least seven cases (3-04A, 3-10A, 3-57A, 3-65A, 3-78C, 3-83A, 3-89A). Dynamical perturbations of the rotation curves discussed here prove that 3-10A and 3-65A are interacting/merger systems. Interactions also involve the 3-78C2 galaxy, and disturbed patterns appear in the  rotation curve of 3-83A1; however, 3-89A1 and 3-89A2 galaxies show unperturbed rotation curves.

We find several systems with previously unsuspected decoupled velocity components, 3-26B, 3-27A, 3-37A, and 3-40A. Moreover peculiar motions arise in the rotation curves of 3-26A, 3-30A, 3-66A, 3-69A, 3-70AA, 3-71A, and 3-76A. Thus,  48\% of the rotation curves have  disturbed morphologies and most part of these are SF galaxies. This emphasizes the role of interactions in
triggering starbursts and, in particular, FIR emission in our sample of dusty galaxies.

\begin{acknowledgements}
We thank the anonymous referee whose comments greatly improved the paper. Some of the data presented herein were obtained at the W.M. Keck Observatory, which is operated as a scientific partnership among the California Institute of Technology, the University of California, and the National Aeronautics and Space Administration. The Observatory was made possible by the generous financial support of the W.M. Keck Foundation.
\end{acknowledgements}

\clearpage
\onecolumn
\begin{landscape}
\begin{longtable}{lccccccccc}
\caption{Fluxes of the principal emission lines: red spectral region} \\
\hline \hline \hline
          & [SII]                      & [SII] & [NII] & H$\alpha$ &  [NII] & [OI] & [OIII]& [OIII] & $A_v$\\
Object    & 6731                      & 6717 & 6583 & 6562    &  6548 & 6300 & 5007 & 4959  & mag\\
\hline
\endfirsthead
\caption{continued}\\
\hline \hline
          & [SII]                      & [SII] & [NII] & H${\alpha}$  & [NII] & [OI] & [OIII]& [OIII] & $A_v$\\
Object    & 6731                      & 6717 & 6583 & 6562    & 6548 & 6300 & 5007 & 4959  & mag\\
\hline
\endhead
3-04A1    & 0.810$\pm$0.477  &  0.345$\pm$0.189   &  2.436$\pm$0.366  & 4.114$\pm$0.311   & 1.192$\pm$0.355 & ---              &    ---               &     ---        &  3.660    \\
3-04A2    & \multicolumn{2}{c}{1.270$\pm$0.178}   &  0.735$\pm$0.211  & 4.196$\pm$0.333   &  ---            & ---              &    ---               &     ---        &  ---   \\
3-10A1    & 2.950$\pm$0.242  &  3.92 $\pm$0.209   &  9.409$\pm$0.286  &20.497$\pm$0.184   & 3.138$\pm$0.194 & 0.779$\pm$0.142  &  2.123$\pm$0.116   &   0.774$\pm$0.234&  3.462         \\
3-10A2    & 1.197$\pm$0.084  &  1.823$\pm$0.34    &  3.466$\pm$0.144  & 8.137$\pm$0.182   & 1.084$\pm$0.114 & 0.568$\pm$0.066  &  0.706$\pm$0.078   &     ---          &  9.534  \\
3-11A     &      ---        &         ---         &  49.68$\pm$12.0   & 14.93$\pm$6.0     & ---             & ---              &    ---               &     ---        &  ---     \\
3-12A     & 1.037$\pm$0.059  &  1.475$\pm$0.052   &  4.413$\pm$0.153  &11.224$\pm$0.078   & 1.361$\pm$0.1   & 0.17 $\pm$0.025  &  0.302$\pm$0.047   &   0.097$\pm$0.025&  9.247  \\
3-14A     & 0.705$\pm$0.31   &  0.54 $\pm$0.225   &  2.537$\pm$0.353  & 6.686$\pm$0.428   & ---             & ---              &    ---               &     ---        &  ---    \\
3-16A*     & 0.052$\pm$0.011  &  0.080$\pm$0.018   &  0.292$\pm$0.022  & 0.429$\pm$0.021   & ---             & ---              &  0.026$\pm$0.008   &     ---          &  9.241     \\
3-17B     & 1.351$\pm$0.46   &  0.5  $\pm$0.214   &  2.352$\pm$0.182  & 3.925$\pm$0.182   & 0.8  $\pm$0.139 & ---              &    ---               &     ---        &  --- \\
3-20A     & 0.80$\pm$0.072  &  1.083$\pm$0.067    &  0.706$\pm$0.135  & 4.069$\pm$0.077   & 0.218$\pm$0.034 & 0.247$\pm$0.039  &  2.904$\pm$0.200   &   1.034$\pm$0.034&  0.911  \\
3-21A     & \multicolumn{2}{c}{3.102$\pm$0.289}   &  1.687$\pm$0.289  & 7.963$\pm$0.322   & ---             &  ---             &  1.298$\pm$0.500   &     ---          &  2.834          \\
3-23A     &   ---           &       ---           &  0.919$\pm$0.122  & 2.169$\pm$0.133   & 0.259$\pm$0.108 & ---              &    ---               &     ---        &  ---   \\
3-24A     & 1.012$\pm$0.106  &  1.254$\pm$0.067   &  4.103$\pm$0.138  & 8.678$\pm$0.234   & 1.359$\pm$0.106 & ---              &  0.451$\pm$0.045   &     ---          &  8.935      \\
3-26A*     & 0.415$\pm$0.035  &  0.567$\pm$0.034   &  0.942$\pm$0.037  & 2.348$\pm$0.037   & 0.294$\pm$0.024 & 0.299$\pm$0.063  &  0.268$\pm$0.030   &     ---          &  6.519       \\
3-26B*     & 0.192$\pm$0.021  &  0.241$\pm$0.020   &  0.519$\pm$0.025  & 1.368$\pm$0.019   & 0.159$\pm$0.021 & ---              &  0.317$\pm$0.015   &   0.103$\pm$0.026&  1.882 \\
3-27A     & 1.94$\pm$0.105  &  2.27 $\pm$0.133    &  6.030$\pm$0.169  &12.378$\pm$0.184   & 1.928$\pm$0.202 & 0.605$\pm$0.13   &  1.581$\pm$0.175   &     ---          &  8.700        \\
3-27B     & ---             &          ---        &  0.580$\pm$0.174  & 1.619$\pm$0.151   & ---             &  ---             &    ---               &     ---        &  ---       \\
3-29A     & 0.467$\pm$0.092  &  0.596$\pm$0.076   &  2.111$\pm$0.08   & 1.833$\pm$0.276   & 0.72 $\pm$0.089 & ---              &  0.154$\pm$0.071   &     ---          &  9.741  \\
3-30A     & 0.251$\pm$0.028  &  0.306$\pm$0.022   &  0.699$\pm$0.023  & 1.408$\pm$0.01    & 0.247$\pm$0.019 & ---              &  0.216$\pm$0.011   &   0.078$\pm$0.058&  7.636   \\
3-31A     & ---             &         ---         &  1.489$\pm$0.988  & 6.014$\pm$0.064   & ---             &  ---             &    ---               &     ---        &  ---       \\
3-36A*     & 0.277$\pm$0.074  &  0.396$\pm$0.026   &  0.928$\pm$0.070   & 2.042$\pm$0.149   & 0.301$\pm$0.044 & ---              &  0.089$\pm$0.026   &     ---          &  7.942 \\
3-36B     & ---             &     ---             &      ---          & 3.722$\pm$0.366   & ---             & ---              &    ---               &     ---        &  ---      \\
3-37A     & 2.879$\pm$0.155  &  3.443$\pm$0.236   & 12.769$\pm$0.374  & 18.56$\pm$0.836   & 2.89 $\pm$0.141 & 0.721$\pm$0.113  &  0.764$\pm$0.158   &     ---          &  8.773  \\
3-38A     & ---             &     ---             &  1.760$\pm$0.278  & 4.669$\pm$0.278   & ---             & ---              &    ---               &     ---        &  ---       \\
3-38B     & \multicolumn{2}{c}{8.186$\pm$0.932}   &  5.510$\pm$0.577  & 9.883$\pm$0.377   & ---             & ---              &    ---               &     ---        &  8.322         \\
3-39A1    & ---             &          ---        &  0.941$\pm$0.079  & 2.494$\pm$0.11    & 0.306$\pm$0.047 & ---              &    ---               &     ---        &  8.577 \\
3-40A     & 0.808$\pm$0.177  &  1.254$\pm$0.236   &  3.980$\pm$0.543  & 6.568$\pm$0.348   & 1.764$\pm$0.189 & 1.386$\pm$0.283  &    ---               &     ---        &  ---   \\
3-42A     & \multicolumn{2}{c}{2.167$\pm$0.200}   &  2.936$\pm$0.2    & 4.347$\pm$0.167   & 1.12 $\pm$0.211 & 0.539$\pm$0.099  &  1.216$\pm$0.160   &   0.265$\pm$0.120&  2.920     \\
3-44A1    & \multicolumn{2}{c}{3.172$\pm$5.290}   &  4.150$\pm$0.238  & 1.956$\pm$0.224   & 2.193$\pm$0.426 & ---              &  2.473$\pm$0.277   &   1.004$\pm$0.410&  ---   \\
3-44A2    & 0.143$\pm$0.06   &  0.332$\pm$0.046   &  0.500$\pm$0.114  & 1.606$\pm$0.143   & 0.164$\pm$0.074 & ---              &  0.085$\pm$0.049   &   0.264$\pm$0.084&  9.151  \\
3-45A*     & 0.211$\pm$0.033  &  0.287$\pm$0.032   &  0.646$\pm$0.031   & 1.327$\pm$0.026   & 0.196$\pm$0.026 & 0.091$\pm$0.031   &    ---               &     ---        &  8.919    \\
3-47A     & ---             &           ---       &  0.771$\pm$0.167  & 2.895$\pm$0.155   & ---             & ---              &    ---               &     ---        &  ---          \\
3-49A1    & ---             &            ---      &  1.703$\pm$0.222  & 3.453$\pm$0.222   & 0.503$\pm$0.109 & ---              &  1.587$\pm$0.090   &   0.461$\pm$0.070&  12.633 \\
3-51A     & 0.834$\pm$0.411  &  0.282$\pm$0.133   &        ---        & 4.527$\pm$0.289   & ---             & ---              &    ---              &     ---         &  ---            \\
3-53A1    & 0.779$\pm$0.051  &  0.932$\pm$0.044   &  1.653$\pm$0.064  & 4.501$\pm$0.254   & 0.398$\pm$0.037 & 0.537$\pm$0.099  &  0.913$\pm$0.084   &   0.472$\pm$0.088&  9.672     \\
3-54A1    & 3.556$\pm$1.443  &  3.158$\pm$0.888   &  3.813$\pm$0.389  &21.778$\pm$0.366   & ---             & 0.904$\pm$0.2    &  5.377$\pm$0.380   &   2.015$\pm$0.400&  2.724         \\
3-55A     & \multicolumn{2}{c}{8.370$\pm$5.161}   &  4.836$\pm$0.455  &17.283$\pm$0.344   & ---             & ---              &  1.217$\pm$0.400   &     ---          &  4.220           \\
3-56A     & ---             &            ---      &  0.640$\pm$0.366  & 2.317$\pm$0.455   & ---             & ---              &    ---               &     ---        &  ---          \\
3-57A1    & 2.436$\pm$0.133  &  2.983$\pm$0.1     &  4.049$\pm$0.91   & 8.880$\pm$1.554   & 1.338$\pm$0.444 & 0.983$\pm$0.244  &  1.137$\pm$0.250   &     ---          &  14.044     \\
3-58A     & ---             &           ---       &  1.626$\pm$0.487  &4.0410$\pm$0.418   & 0.51 $\pm$0.371 & ---              &    ---               &     ---        &  --- \\
3-59A     & \multicolumn{2}{c}{4.043$\pm$0.133}   &  4.052$\pm$0.244  & 9.315$\pm$0.266   & 0.737$\pm$0.233 & ---              &    ---               &     ---        &  11.306   \\
3-61A     & 1.331$\pm$0.488  &  1.694$\pm$0.3     &  3.773$\pm$0.466  & 8.116$\pm$0.511   & ---             &  ---             &    ---               &     ---        &  ---         \\
3-64A     & \multicolumn{2}{c}{11.037$\pm$0.800}  &  4.094$\pm$0.777  &26.196$\pm$1.021   & ---             & ---              &    ---               &     ---        &  ---         \\
3-65A1    & 2.543$\pm$0.059  &  3.371$\pm$0.076   &  8.507$\pm$0.446  &22.833$\pm$0.209   & 2.234$\pm$0.446 & 0.431$\pm$0.047  &  1.333$\pm$0.154   &   0.378$\pm$0.034&  7.702    \\
3-65A2    & 0.854$\pm$0.044  &  1.148$\pm$0.103   &  1.522$\pm$0.276  & 5.308$\pm$0.324   & ---             & 0.346$\pm$0.047  &  0.585$\pm$0.036   &   0.147$\pm$0.023&  9.654          \\
3-66A     & 1.119$\pm$0.038  &  1.624$\pm$0.058   &  1.678$\pm$0.077  & 8.994$\pm$0.09    & 0.544$\pm$0.076 & 0.281$\pm$0.077  &  3.541$\pm$0.066   &   1.224$\pm$0.075&  2.247    \\
3-69A*     & 0.172$\pm$0.009  &  0.197$\pm$0.014   &  0.819$\pm$0.026  & 1.391$\pm$0.092   & 0.271$\pm$0.025 & 0.025$\pm$0.016  &  0.072$\pm$0.026   &     -            &  5.912    \\
3-70A     & 0.467$\pm$0.018  &  0.566$\pm$0.068   &  3.037$\pm$0.024  & 2.524$\pm$0.109   & 0.881$\pm$0.021 & 0.185$\pm$0.015  &  4.186$\pm$0.160   &   1.468$\pm$0.181&  4.481    \\
3-71A*     & 0.695$\pm$0.057  &  0.842$\pm$0.058   &  2.230$\pm$0.118  & 5.545$\pm$0.110   & 0.657$\pm$0.041 & 0.124$\pm$0.034  &  1.222$\pm$0.147   &   0.537$\pm$0.135&  13.073     \\
3-73A     & ---             &           ---       &  3.097$\pm$0.666  & 6.593$\pm$0.955   & 0.619$\pm$0.566 & ---              &    ---               &     ---        &  ---  \\
3-75A     & 0.109$\pm$0.078  &  0.193$\pm$0.041   &  0.366$\pm$0.061  & 0.723$\pm$0.058   & 0.092$\pm$0.056 & ---              &    ---               &     ---        &  ---   \\
3-76A*     & 0.150$\pm$0.034  &  0.182$\pm$0.026   &  0.478$\pm$0.053  & 1.172$\pm$0.053   & 0.155$\pm$0.022 & ---              &    ---               &     ---        &  7.257  \\
3-77A     & 0.808$\pm$0.466  &  0.958$\pm$0.477   &  1.524$\pm$0.155  & 3.629$\pm$0.333   & 0.491$\pm$0.721 & ---              &    ---               &     ---        &  ---    \\
3-78A     & ---             &            ---      &  0.719$\pm$0.144  & 2.298$\pm$0.155   & ---             & ---              &    ---               &     ---        &  ---          \\
3-78B     & \multicolumn{2}{c}{3.064$\pm$0.220}   &  1.871$\pm$0.133  & 6.219$\pm$0.167   & 0.575$\pm$0.178 & ---              &    ---               &     ---        &  4.394   \\
3-78C1    & 0.135$\pm$0.009  &  0.175$\pm$0.025   &  0.383$\pm$0.027  & 1.067$\pm$0.034   & 0.144$\pm$0.018 & ---              &  0.114$\pm$0.011   &   0.039$\pm$0.015&  6.917       \\
3-78C2    & 0.137$\pm$0.027  &  0.181$\pm$0.018   &  0.385$\pm$0.045  & 1.069$\pm$0.038   & 0.127$\pm$0.024 & ---              &  0.115$\pm$0.013   &   0.039$\pm$0.015&  6.922      \\
3-79A     & 4.913$\pm$1.332  &  7.027$\pm$1.221   &  19.86$\pm$0.966  &41.248$\pm$0.844   & 5.843$\pm$1.11  & 0.472$\pm$0.322  &    ---               &     ---        &  4.465      \\
3-79C     & \multicolumn{2}{c}{3.213$\pm$0.300}   &  5.017$\pm$0.311  & 9.105$\pm$0.377   & 1.616$\pm$0.4   & ---              &  2.378$\pm$0.390   &   0.482$\pm$0.320&  4.102   \\
3-81A     & ---             &     ---             &  8.790$\pm$0.844  &23.843$\pm$1.11    & ---             & ---              &    ---             &     ---          &  ---        \\
3-81B     &\multicolumn{2}{c}{106.016$\pm$6.438}  &127.317$\pm$4.551 &530.025$\pm$5.772   & 35.92$\pm$7.77  & 5.289$\pm$3.885  & 182.40$\pm$9.700   &     ---          &  ---      \\
3-83A1    & 0.071$\pm$0.014  &  0.093$\pm$0.021   &  0.126$\pm$0.027  & 0.352$\pm$0.058   & ---             & ---              &  0.056$\pm$0.028   &     ---          &  5.625         \\
3-83A2    & 0.417$\pm$0.013  &  0.549$\pm$0.018   &  1.102$\pm$0.034  & 5.141$\pm$0.133   & 0.378$\pm$0.029 & 0.066$\pm$0.012  &  0.880$\pm$0.021   &   0.322$\pm$0.024&  5.574     \\
3-84A1    & 4.548$\pm$0.433  &  2.7  $\pm$0.255   &  5.498$\pm$0.133  & 15.54$\pm$0.108   & 1.679$\pm$0.107 & 1.618$\pm$0.092  &  8.684$\pm$0.150   &   2.759$\pm$0.130&  5.266    \\
3-85A1    & ---             &            ---      &  1.282$\pm$0.266  & 1.726$\pm$0.189   & ---             & ---              &    ---             &     ---          &  5.920       \\
3-88A1    & 4.509$\pm$1.332  &  7.087$\pm$1.01    &  2.808$\pm$0.644  &51.204$\pm$0.699   & ---             & 1.815$\pm$0.3    & 55.480$\pm$1.000   &  16.500$\pm$0.950&  1.151         \\
3-89A1    & 0.516$\pm$0.034  &  0.781$\pm$0.045   &  0.985$\pm$0.039  & 4.794$\pm$0.223   & 0.3  $\pm$0.038 & 0.143$\pm$0.049  &  1.208$\pm$0.219   &   0.465$\pm$0.034&  6.393    \\
3-89A2    & 0.298$\pm$0.087  &  0.479$\pm$0.099   &  1.061$\pm$0.115  & 2.284$\pm$0.107   & 0.347$\pm$0.078 & ---              &    ---             &     ---          &  ---   \\
3-90A     & 0.457$\pm$0.043  &  0.509$\pm$0.038   &  1.413$\pm$0.051  & 3.407$\pm$0.061   & 0.456$\pm$0.036 & 0.041$\pm$0.016  &  0.128$\pm$0.040   &     ---          &  7.283    \\
3-91A     & \multicolumn{2}{c}{6.160$\pm$0.488}   &  3.849$\pm$0.344  &18.071$\pm$0.333   & ---             & ---              &  3.696$\pm$0.430   &   1.114$\pm$0.370&  3.437   \\
3-92A     & 1.487$\pm$0.111  &  2.03 $\pm$0.093   &  2.415$\pm$0.311  &13.353$\pm$0.322   & ---             & 0.768$\pm$0.133  &  3.258$\pm$0.350   & 1.047$\pm$0.220  &  8.252            \\
3-92C     & \multicolumn{2}{c}{4.359$\pm$0.455}   &  2.922$\pm$0.4    &12.321$\pm$0.333   & ---             & ---              &  4.230$\pm$0.200   &   1.526$\pm$0.290&  7.613          \\
3-93A     & \multicolumn{2}{c}{3.746$\pm$0.266}   &  1.952$\pm$0.2    & 9.398$\pm$0.189   & ---             & 0.199$\pm$0.133  &  2.220$\pm$0.320   &   0.966$\pm$0.340&  3.654          \\
3-96A     & 0.266$\pm$0.029  &  0.232$\pm$0.026   &  1.045$\pm$0.019  & 0.748$\pm$0.027   & 0.303$\pm$0.016 & 0.145$\pm$0.015  &  2.397$\pm$0.042   &   0.884$\pm$0.043&  5.359      \\
\hline
\label{tab_em}
\end{longtable}
The fluxes, in unit of 10$^{-18}$ erg s$^{-1}$cm$^{-2}\,\AA^{-1}$, are not corrected for internal absorption. For H$\alpha$  we report here only the narrow component. The fluxes of the [SII] doublet are given separately, if available, otherwise the total flux of the doublet is given in the middle of the columns of the single lines.
\end{landscape}
\begin{longtable}{lcccc}
\caption{Fluxes of the principal emission lines: blue spectral region } \\
\hline \hline
         & H$_{\beta}$  & H$_{\gamma}$  & [OII]                       & [OII] \\
Object    &  4861 & 4340 & 3729                       & 3726 \\
\hline
\endfirsthead
\caption{continued}\\
\hline \hline \hline
          &H$_{\beta}$  & H$_{\gamma}$  & [OII]                       & [OII] \\
Object    & 4861 & 4340 & 3729                       & 3726 \\
\hline \hline
\endhead
3-04A1    &   0.953$\pm$0.104  & ---  & ---                        & ---  \\
3-04A2    &  ---    & ---  & ---                        & ---  \\
3-10A1    &   4.854$\pm$0.113  & 1.756$\pm$0.087 & \multicolumn{2}{c}{7.278$\pm$1.357}          \\
3-10A2    &   0.973$\pm$0.156  & 0.246$\pm$0.046 & 0.849$\pm$0.328               & 0.769$\pm$0.265 \\
3-11A     &   ---    & ---  & ---                        & ---  \\
3-12A     &   1.386$\pm$0.126  & 0.414$\pm$0.028 & 0.557$\pm$0.138                       & 0.705$\pm$0.318 \\
3-14A     &   ---    & ---  & ---                        & ---  \\
3-16A*     &   0.053$\pm$0.021  & ---  & ---                        & ---  \\
3-17B     &   ---    & ---  & ---                        & ---  \\
3-20A     &   1.284$\pm$0.051  & 0.656$\pm$0.033 & 3.477$\pm$0.191                       & 2.362$\pm$0.191 \\
3-21A     &   2.024$\pm$0.364  & ---  & ---                        & ---  \\
3-23A     &   ---    & ---  & ---                        & ---  \\
3-24A     &   1.11 $\pm$0.155  & 0.229$\pm$0.059 & \multicolumn{2}{c}{1.592$\pm$0.857}           \\
3-26A*     &   0.394$\pm$0.051  & 0.082$\pm$0.021 & \multicolumn{2}{c}{0.863$\pm$0.191}          \\
3-26B*     &   0.387$\pm$0.110  & 0.08$\pm$0.015 & ---                        & ---  \\
3-27A     &   1.625$\pm$0.101  & 0.519$\pm$0.058 & \multicolumn{2}{c}{1.987$\pm$0.574}          \\
3-27B     &   ---    & ---  & ---                        & ---  \\
3-29A     &   0.214$\pm$0.105  & ---  & ---                        & ---  \\
3-30A     &   0.208$\pm$0.076  & 0.069$\pm$0.013 & \multicolumn{2}{c}{0.702$\pm$0.077}          \\
3-31A     &   ---    & ---  & ---                        & ---  \\
3-36A*     &   0.292$\pm$0.032  & 0.085$\pm$0.037 & ---                        & ---  \\
3-36B     &   ---    & ---  & ---                        & ---  \\
3-37A     &   2.417$\pm$00.737 & 0.969$\pm$0.258 & ---                        & ---  \\
3-38A     &   ---    & ---  & ---                        & ---  \\
3-38B     &   1.354$\pm$0.322  & ---  & ---                        & ---  \\
3-39A1    &   0.333$\pm$0.097  & ---  & ---                        & ---  \\
3-40A     &   ---    & ---  & ---                        & ---  \\
3-42A     &   1.094$\pm$0.156  & ---  & ---                        & ---  \\
3-44A1    &   ---    & ---  & ---                        & ---  \\
3-44A2    &   0.2  $\pm$0.036  & 0.047$\pm$0.031 & 0.234$\pm$0.064           & 0.105$\pm$0.050 \\
3-45A*     &   0.170 $\pm$0.046  & ---  & ---                        & ---  \\
3-47A     &   ---    & ---  & ---                        & ---  \\
3-49A1    &   0.291$\pm$0.076  & ---  & ---                        & ---  \\
3-51A     &   ---    & ---  & ---                        & ---  \\
3-53A1    &   0.53 $\pm$0.019  & 0.152$\pm$0.017 & \multicolumn{2}{c}{1.851$\pm$0.024}          \\
3-54A1    &   5.604$\pm$0.499  & ---  & ---                        & ---  \\
3-55A     &   3.758$\pm$0.51   & ---  & ---                        & ---  \\
3-56A     &   ---    & ---  & ---                        & ---  \\
3-57A1    &   0.64 $\pm$0.218  & ---  & ---                        & ---  \\
3-58A     &   ---    & ---  & ---                        & ---  \\
3-59A     &   0.912$\pm$0.302  & ---  & ---                        & ---  \\
3-61A     &   ---    & ---  & ---                        & ---  \\
3-64A     &   ---    & ---  & ---                        & ---  \\
3-65A1    &   3.354$\pm$0.049  & 1.128$\pm$0.044 & 2.361$\pm$0.095                     & 2.865$\pm$0.092 \\
3-65A2    &   0.626$\pm$0.072  & 0.160$\pm$0.01 & 0.450$\pm$0.088        & 0.779$\pm$0.127 \\
3-66A     &   2.442$\pm$0.066  & 0.523$\pm$0.106 & ---                        & ---  \\
3-69A*     &   0.250$\pm$0.055  & 0.120$\pm$0.072 & ---                        & ---  \\
3-70A     &   0.533$\pm$0.035  & 0.158$\pm$0.022 & 0.396$\pm$0.024      & 0.296$\pm$0.027 \\
3-71A*    &   0.445$\pm$0.087  & 0.072$\pm$0.024 & ---                        & ---  \\
3-73A     &   ---    & ---  & ---                        & ---  \\
3-75A     &   ---    & ---  & ---                        & ---  \\
3-76A*     &   0.181$\pm$0.054  & ---  & ---                        & ---  \\
3-77A     &   ---    & ---  & ---                        & ---  \\
3-78A     &   ---    & ---  & ---                        & ---  \\
3-78B     &   1.326$\pm$0.229  & ---  & ---                        & ---  \\
3-78C1    &   0.171$\pm$0.015  & 0.063$\pm$0.014 & \multicolumn{2}{c}{0.287$\pm$0.020}          \\
3-78C2    &   0.17 $\pm$0.017  & 0.063$\pm$0.019 & \multicolumn{2}{c}{0.290$\pm$0.024}          \\
3-79A     &   8.725$\pm$0.77   & ---  & ---                        & ---  \\
3-79C     &   2.006$\pm$0.25   & ---  & ---                        & ---  \\
3-81A     &   ---    & ---  & ---                        & ---  \\
3-81B     &   ---    & ---  & ---                        & ---  \\
3-83A1    &   0.065$\pm$0.03   & ---  & ---                        & ---  \\
3-83A2    &   0.96 $\pm$0.035  & 0.258$\pm$0.017 & 0.849$\pm$0.049    & 0.606$\pm$0.057 \\
3-84A1    &   3.017$\pm$0.135  & ---  & ---                        & ---  \\
3-85A1    &   0.31 $\pm$0.125  & ---  & ---                        & ---  \\
3-88A1    &  14.373$\pm$1.04  & ---  & ---                        & ---  \\
3-89A1    &   0.816$\pm$0.204 & 0.232$\pm$0.049 & 0.864$\pm$0.061     & 0.716$\pm$0.084 \\
3-89A2    &   ---    & ---  & ---                        & ---  \\
3-90A     &   0.525$\pm$0.031  & ---  & ---                        & ---  \\
3-91A     &   4.291$\pm$0.312  & ---  & ---                        & ---  \\
3-92A     &   1.844$\pm$0.302  & ---  & ---                        & ---  \\
3-92C     &   1.828$\pm$0.218  & ---  & ---                        & ---  \\
3-93A     &   2.178$\pm$0.291  & ---  & ---                        & ---  \\
3-96A     &   0.143$\pm$0.01   & 0.032$\pm$0.014 & \multicolumn{2}{c}{0.508$\pm$0.024}          \\
\hline
\label{tab_em1}
\end{longtable}
The fluxes are in unit of 10$^{-18}$ erg s$^{-1}$cm$^{-2}\,\AA^{-1}$. The fluxes  of the [OII] doublet are given separately, if available, otherwise the total flux of the doublet is given in the middle of the columns of the single lines.

\begin{longtable}{llllll}
\caption{Spectral classification} \\
\hline \hline
Name  & Class. & Name  & Class. & Name  & Class. \\
\hline
\endfirsthead
\caption{continued}\\
\hline \hline
Name  & Class. & Name  & Class. & Name  & Class. \\
\hline
\endhead
3-10A1 & SF      & 3-42A  & L     & 3-78C1 & SF   \\ 
3-10A2 & SF      & 3-44A1 & Sy   & 3-78C2 & SF   \\ 
3-12A  & SF      & 3-44A2 & SF   & 3-79C  & SF   \\ 
3-16A  & SF      & 3-49A1 & L     & 3-83A1 & SF   \\ 
3-20A  & L        & 3-53A1 & L    & 3-83A2 &  SF    \\ 
3-21A  & SF      & 3-54A1 & SF    & 3-84A1 & L   \\ 
3-24A  & SF      & 3-55A  & SF    & 3-88A1 & SF  \\ 
3-26A  & L        & 3-57A1 & L      & 3-89A1 & SF     \\ 
3-26B  & SF      & 3-65A1 & SF    & 3-90A  & SF     \\ 
3-27A  & SF      & 3-65A2 & SF    & 3-91A  & SF   \\ 
3-29A  & L        & 3-66A  & SF    & 3-92A  & SF     \\ 
3-30A  & SF      & 3-69A  & SF    & 3-92C  & L \\ 
3-36A  & SF      & 3-70A  & Sy    & 3-93A  & SF     \\ 
3-37A  & SF      & 3-71A  & SF    & 3-96A  & Sy    \\ 
\hline\hline
\label{tab_sp_class}
\end{longtable}
\newpage
\begin{longtable}{lcccccccc|cc|c}
\caption{Kinematical parameters and masses. } \\
\hline \hline
Name & $z$ & $i$   & $\delta$ & $r_{\mathrm opt}$ & $R_{\mathrm opt}$ &
$v_{\mathrm opt}$      & $v_c$& Mass   & $r_t$\\
     &     & $[~^\circ~]$ & $[~^\circ~]$ & $['']$ & [Kpc]  &
[km/s]  & [km/s]& [$\log M_\odot$] & $['']$\\
\hline
\endfirsthead
\caption{continued}\\
\hline \hline
Name & $z$ & $i$   & $\delta$ & $r_{\mathrm d}$ & $R_{\mathrm d}$ &
$v_{\mathrm opt}$   & $v_c$&  Mass     & $r_t$\\
     &     & $[~^\circ~]$ & $[~^\circ~]$ & $['']$ & [Kpc]  &
[km/s]   & [km/s]&  [$\log M_\odot$]&$['']$\\
\hline
\endhead
3-10A1 & 0.0875 & 48 & 22 &  5.99 & 9.79 & 100   & 133& 11.65 & 0.6\\
3-10A2 & 0.0867 & 38 & 74 &  2.99 & 4.50 &  50   & 271& 11.92 & 0.4 \\
3-12A & 0.0775 & 24 & 14 &  5.97 & 8.74 &  25  & 58&  10.89 & --- \\
3-16A & 0.1176 & 51 &  6 &  3.31 & 7.33 & 117   & 140& 11.57 & 1.8\\
3-20A & 0.0736 & 71 & 15 &  6.78 & 9.43 & 90 &  92& 11.31  & --- \\
3-23A$^{*}$ & 0.0877 & 66 & 37 &  1.38 &  2.26 &  30  & 38 &  9.91 & --- \\
3-24A & 0.1164 & 26 & 12 &  3.13 & 6.59 & 110   & 230& 11.95 & 0.4 \\
3-26A & 0.0894 & 35 & 13 &  4.88 & 8.12 & 145  & 238& 12.07  & 0.8 \\
3-26B & 0.0889 & 73 &  4 &  9.50 & 15.75 & 170   & 164& 12.03 & 7.5\\
3-27A & 0.0876 & 52 & 39 &  6.87 &  11.24 & 100   & 150& 11.81 & 3.8\\
3-29A & 0.0408 & 61 & 24 &  3.21 &  2.58 & 117  & 141 & 11.11 & 1.5\\
3-30A & 0.2545 & 51 & 12 &  2.05 & 8.12&  59 &  62 & 10.89 & 0.2 \\
3-31A$^{*}$ & 0.0267 & 37 &  5 & 16.50 &  8.79 & 50  & 81& 11.17 &---\\
3-36A & 0.1183 & 69 &  0 &  3.01 &  6.43 & 145  & 139& 11.50 & 0.4 \\
3-37A & 0.0507 & 43 & 16 &  3.06 &  3.02 &  30 &  44& 10.16  & 0.4 \\
3-38A$^{*}$ & 0.0883 & 40 & 25 &  2.75 &  4.53 & 45  & 71& 10.76  & ---\\
3-40A & 0.0892 & 67 & 11 &  2.39 & 3.82 & 120 & 123&  11.16  & 0.8 \\
3-44A2 & 0.3027 & 52 & 58 &  1.44 & 6.46 & 95  & 175&  11.70 & 0.5 \\
3-45A & 0.0791 & 64 &  2 &  4.28 &  6.39 & 176  & 182 & 11.73& 1.3 \\
3-47A$^{*}$ & 0.0874 & 68 & 42 &  1.38 &  2.25 & 228  & 304&  11.72 & ---\\
3-53A1 & 0.3723 & 22 & 25 & 0.74 &  3.80 & 36 & 77&  10.76 & 0.1 \\
3-59A$^{*}$ & 0.0865 & 34 & 28 &  5.23 & 8.47 & 160  & 298&  12.28 & --\\
3-61A$^{*}$ & 0.0545 & 50 & 47 &  6.33 &  6.69 & 306& 555 &  12.72  & ---\\
3-65A1 & 0.1667 & 39 & 33 &  2.00 &  5.70 &  32   &  52&  10.59 & 0.4 \\
3-65A2 & 0.1672 & 33 & 88 &  2.47 &  6.49 &  11 &   ---   &  --- & 0.1\\
3-66A & 0.0536 & 62 &  5 &  7.50 &  7.80 &  90  &  97&   11.27 & 0.9 \\
3-69A & 0.1043 & 47 &  2 &  5.21 &  9.96 &  29  &  36 &   10.51& 1.3 \\
3-70A & 0.1970 & 17 & 78 &  1.65 &  5.38 &  80 &  ---   & ---   & --- \\
3-71A & 0.0516 & 42 &  8 &  4.28 &  4.29 & 110 & 161 &  11.45   & ---  \\
3-76A & 0.0800 & 33 & 58 &  5.05 &  7.61 & 35  & 112&   11.39 & 0.2 \\
3-78C1 & 0.2610 & 69 & 19 &  2.04 &  8.23 & 153 & 138&  11.60  & 0.6 \\
3-78C2 & 0.2611 & 83 & 50 &  1.76 &  7.10 &  25  &  31&  10.24 & 0.2 \\
3-81A$^{*}$ & 0.0282 & 42 & 74 &  4.95 &  2.78 & 45  & 237 &  11.60& --- \\
3-81B$^{*}$ & 0.0264 & 85 & 30 &  1.00 & 0.56  & 90   & 102& 10.17   & --- \\
3-83A1 & 0.1071 & 61 &  5 &  4.70 & 9.20 & 190  & 197&  11.96 & 2.4 \\
3-83A2 & 0.1071 & 35 & 12 &  2.85 & 5.78 & 55  &  89&   10.06 & 1.7 \\
3-88A1$^{*}$ & 0.0516 & 67 & 32 &  2.48 &  2.46 & 180 & 219&  11.48  & --- \\
3-89A1 & 0.2996 & 33 & 52 &  1.54 & 6.84 &  39  &  90&   11.14 & 0.3 \\
3-89A2 & 0.2992 & 46 & 0 &  1.00 & 4.45 & 160& 171  &  11.52 & 0.2 \\
3-90A & 0.0721 & 42 & 15 &  4.10 &  5.62 & 118 & 170&  11.62  & 0.8 \\
3-93A$^{*}$ & 0.0691 & 74 & 10 &  4.40 &  5.80 &  16  & 16&   9.57 & ---\\
\hline
\label{tab_rotcur}
\end{longtable}
\small{
Columns: (1)  galaxy name, (2) redshift, (3) galaxy inclination, (4) misalignment angle, (5) and (6) maximum radius of the rotation curve in arcsec and Kpc, respectively, (7) rotation velocity measured at R$_{opt}$, (8) mass derived by equation (1), (9) circular velocity, i.e., rotation velocity accounting for inclination and misalignment corrections (10)  transition radius between the rising and flat parts of the rotation curve; $*$ galaxies with TNG spectra.\\
}
\end{document}